\documentclass[sigconf, nonacm]{acmart}

\usepackage{graphicx}
\usepackage{amsmath}
\usepackage{booktabs}
\usepackage{epigraph}
\usepackage{algorithm}
\usepackage{algpseudocode}
\usepackage{comment}
\usepackage{enumitem}
\setlist{nolistsep}
\usepackage{dblfloatfix}
\usepackage{subcaption}
\usepackage{array}
\usepackage{multirow}
\AtBeginDocument{%
  }

\begin{document}

\title{"I don't like my avatar": Investigating Human Digital Doubles}

\author{Siyi Liu}
\affiliation{%
  \institution{Utrecht University}
  \city{Utrecht}
  \country{Netherlands}}
\email{1312156cao@gmail.com}

\author{Kazi Injamamul Haque}
\affiliation{%
  \institution{Utrecht University}
  \city{Utrecht}
  \country{Netherlands}}
\email{k.i.haque@uu.nl}

\author{Zerrin Yumak}
\affiliation{%
  \institution{Utrecht University}
  \city{Utrecht}
  \country{Netherlands}}
\email{z.yumak@uu.nl}

\renewcommand{\shortauthors}{Liu, Haque and Yumak}


\begin{abstract}
Creating human digital doubles is becoming easier and much more accessible to everyone using consumer grade devices. In this work, we investigate how avatar style (realistic vs cartoon) and avatar familiarity (self, acquaintance, unknown person) affect self/other-identification, perceived realism, affinity and social presence with a controlled offline experiment. We created two styles of avatars (realistic-looking MetaHumans and cartoon-looking ReadyPlayerMe avatars) and facial animations stimuli for them using performance capture. Questionnaire responses demonstrate that higher appearance realism leads to a higher level of identification, perceived realism and social presence. However, avatars with familiar faces, especially those with high appearance realism, lead to a lower level of identification, perceived realism, and affinity. Although participants identified their digital doubles as their own, they consistently did not like their avatars, especially of realistic appearance. But they were less critical and more forgiving about their acquaintance's or an unknown person's digital double.
\end{abstract}


\begin{teaserfigure}
\centering
  \includegraphics[width=0.95\textwidth,trim={5mm 18mm 5mm 5mm},clip]{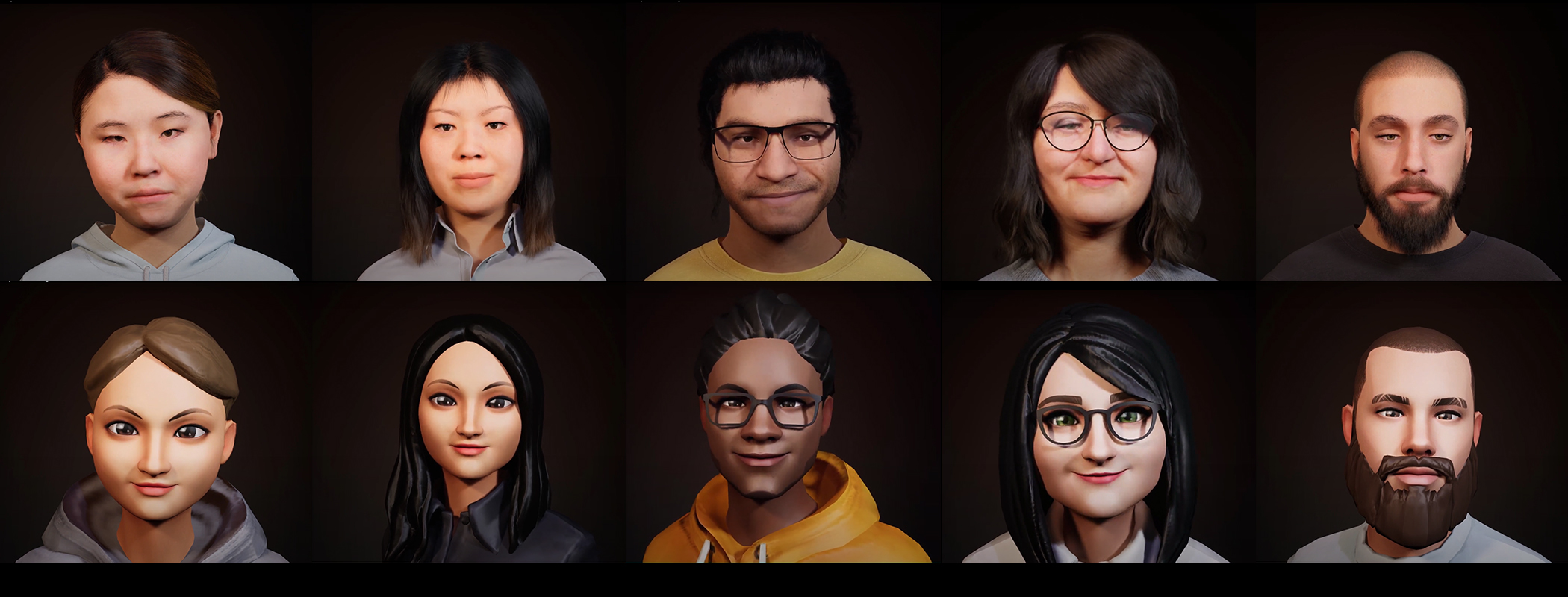}
  \caption{We investigate the perception of human digital doubles with realistic (top row) and stylized (bottom row) appearances and three levels of familiarity - self, acquaintance and unknown person.}
  \Description{Teaser figure containing examples of human digital doubles.}
  \label{fig:teaser}
\end{teaserfigure}

\maketitle

\section{Introduction}
We have seen game creators avidly using realistic digital doubles of actors/actresses in video games in recent years. Keanu Reeves’ digital double Johnny Silverhand appears in \textit{Cyberpunk 2077} while Gina Torres appears as Kirkan in \textit{Immortals of the Aveum}- to name a few. While gamers and creators from around the world were amazed by the visual results of such characters, we raised a question- \textit{Did Keanu and Gina like their own digital doubles as they appeared in the respective games?} This question is more relevant when the rapid advancements in computer graphics and artificial intelligence are dramatically transforming the accessibility of creating human digital doubles and how they are used, commonly known as avatars. No longer confined to the realm of specialized, high-cost equipment, these digital representations can now be generated using consumer-grade devices, making them more widely available to the general people. This democratization of digital double creation is not just revolutionizing entertainment industries, such as movies and video games, but is also fueling the burgeoning interest in the "metaverse"— a concept that envisions immersive, interconnected virtual worlds where individuals interact through their avatars \cite{2022metaverse}. 

Given the current expansion of virtual avatar applications beyond video games, into various fields such as remote collaboration \cite{nicetomeetyou, higgins2021remotely}, education \cite{avataronlineeducation, avatarethicaleducation}, virtual socialization \cite{vrchat}, and healthcare \cite{Avatarswithfacesofrealpeople, avatarinpatienteducation}, understanding user perception of these avatars is becoming increasingly important. As spatial computing and extended reality (XR) technologies continue to evolve, the potential applications of human digital doubles are broadening too, prompting a surge in not only academic endeavors but also in industry. Researchers are now focusing on both the psychological implications of these avatars as well as how users perceive them in various contexts.

Previous avatar perception studies have primarily explored how varying levels of visual realism influence user perception, oftentimes using generic digital avatars that are not designed following the appearance of an existing human being \cite{humanorrobot, Tinwell2013, photorealismavoidtheuncanny, NabilaAmadou}. A number of studies investigated  personalized digital avatars, digital doubles of existing people e.g. using photogrammetry but they usually fall short in terms of high-fidelity \cite{higgins2021remotely, Kim2023, VisualFidelity} mainly due to access to high-end systems. Some studies investigated perception of stylized vs. realistic digital humans, such as- Fraser et al. \cite{FRASER2024100082}, whilst some investigated the perception of generic vs. personalized avatars, such as- Salagean et al. \cite{VirtualTwin}. However, familiarity of the avatars is not investigated well. The extent to which an avatar resembles a familiar face— whether that of oneself, an acquaintance, or a stranger (analogous to generic avatars)— could significantly impact user perception. Although this familiarity aspect has been studied in the context of neural responses of participants for face recognition task of real human faces \cite{Familiarity_CAMPBELL2020107415}, familiarity investigation in the context of avatar perception with realistic and stylized representations, remains underexplored. Insights gained from such research are vital for optimizing the use of human avatars in diverse applications related to education, remote collaboration, social XR, healthcare etc. 

In this study, we present a novel investigation into how avatar style (i.e., realistic vs. stylized) in conjunction with familiarity (i.e. self, acquaintance, unknown person) influence self/other identification, perceived realism, affinity, and social presence. Avatars are personal digital representations in a virtual 3D environment that can either have direct resemblance with a real-life person\cite{VirtualTwin, FRASER2024100082} or can also appear as someone entirely different, oftentimes can be stylized characters or even animals \cite{Kyrlitsias2022}. In our study, we focus on the former- avatars having human resemblance with real-life, existing people. We conceptualize this category of human avatars as "Digital Doubles" as these specific human avatars are digital representations of the users having visual resemblance. In our study, we follow Fraser et al.\cite{FRASER2024100082} and use two versions of digital doubles in terms of appearance: stylized avatar (low human resemblance with high graphical resolution) and realistic avatar (high human resemblance with high graphical resolution).

Unlike previous research that suggested a general preference for realistically rendered digital humans \cite{nicetomeetyou, photorealismavoidtheuncanny, IsPhotorealismImportant,NabilaAmadou}, our study specifically examines how these preferences shift when the digital double bears a familiar face. Our findings reveal that while higher appearance realism generally enhances self/other identification, perceived realism, and social presence, familiar avatars, especially those with high realism, paradoxically lead to lower levels of self/other-identification, perceived realism, and affinity. Interestingly, participants were consistently more critical of their own digital doubles, particularly when these had realistic style, but displayed greater tolerance and acceptance towards digital doubles of acquaintances or unknown individuals. 

\section{Related Work}
Previous related works pertaining to digital/virtual humans investigated both perceptual and psychological effects on users. Perceptual studies not only investigated the visual perception \cite{IsPhotorealismImportant, RenderMeReal, FacialFeature}, but also auditory \cite{mitchell2011mismatch, voice2, voice3}. The other line of works looked into psychological aspects of digital humans including- trust \cite{trust, trust2}, personality \cite{PersonalityPan2015, Personality2, personality3}, empathy \cite{EmpathyHiggins2022, empathicresponses} and human collaboration \cite{humancollab}. Our study focuses on the visual perception of virtual humans in terms of self/other identification, realism, affinity and social presence with an emphasis on digital doubles of real humans with varied familiarity.

\subsection{Perception of Virtual Humans and Personalized Avatars}
\label{sec:perceptionofvh}
Numerous studies have demonstrated the significant impact of avatars on users’ perceptions in virtual experiences. User perception encompasses multiple dimensions, including the avatar’s appearance, animation and interaction with the users as well the context/environment. The realism of avatars and their fidelity to real-world appearance are critical in shaping and enhancing virtual experiences. Previous work looked into dimensions such as self-identification, perceived visual realism, affinity and social presence as a measure of perceptual quality. In the next subsections, we explain these concepts and previous work related with these. 

\subsubsection{Self-identification}: Numerous studies have explored individuals' \textit{sense of self-identification} with avatars representing themselves \cite{VirtualTwin, Mal2024_oddone, VisualFidelity, higgins2021remotely}. Kang et al.~\cite{slefidentification} show that associating an experience with oneself, especially from a first-person perspective, enhances self-identification. They define self-identification as the extent to which one feels connected to or represented by their avatar, based on shared attributes such as appearance and behavior. High self-identification occurs when users perceive strong overlap between themselves and the avatar. Gonzalez-Franco et al.~\cite{usingfacialanimation} emphasize the importance of facial appearance and animation in identity construction, showing that facial cues are critical for recognition. Even in first-person applications, users may develop a sense of self-identity through avatar facial features, a phenomenon known as the "enfacement illusion." This refers to the perceptual experience of adopting a virtual or artificial face as one's own. Prior studies have demonstrated this effect using self-other face recognition tasks \cite{sforza2010myface, tajadura2012otherme, tajadura2012personmirror}, where participants view a morphing sequence between self and avatar faces \cite{facetask, usingfacialanimation, VirtualTwin}. Gonzalez-Franco et al. \cite{usingfacialanimation} proposed that embodying an avatar in front of a mirror can amplify the enfacement effect. Salagean et al.~\cite{VirtualTwin} found that avatars with high photorealism and personalization enhance the enfacement illusion. 
The identification of avatars representing self as well as others is the focus of our study. Recognizing interaction partners as real individuals can enhance user engagement in virtual experiences. For instance, Matthew et al. \cite{Avatarswithfacesofrealpeople} asked participants to identify realistic avatars by matching them with real-life photos. The study showed that avatars created from 2D facial photos retained sufficient recognizable features for successful identification. We argue that self-identification concept presented in Kang et al. \cite{slefidentification} can be generalized to others' avatar identification by representing appearance, behavior and personalized features of an existing human being with a digital double. This existing human can be familiar to a certain person or can be unfamiliar/stranger. The studies mentioned above \cite{VirtualTwin, resize} have explored self-identity perception and self-presence in scenes where individuals view their virtual avatars in a virtual mirror. However, they did not investigate how viewers perceive the identity of communication partners' avatars during interactions. Moreover, the fidelity of avatar representations were not high. Our work aims to address this gap by utilizing high fidelity realistic digital humans and investigates both self and others' identification using varying levels of familiarity. The concept of familiarity \cite{rheingold1985development} is defined in psychology and studied especially in the context of trust in interaction with virtual humans \cite{Song17012024}. However, it was not studied in the context of digital doubles and identity perception. In this study, we use self-identification concept as defined in \cite{slefidentification} and develop our own questionnaire items to extend self- identification to other-identification. For familiarity, we adopt the categories from Campbell et al. \cite{Familiarity_CAMPBELL2020107415}: the own-face (high familiarity), a friend's face (moderate familiarity), and a stranger's face (no familiarity). 

\subsubsection{Perceived visual realism}: Research on \textit{perceived visual realism} of virtual avatars spans multiple dimensions, including shape style \cite{zell2015stylize, VisualFidelity, nicetomeetyou}, rendering style \cite{IsPhotorealismImportant, RenderMeReal, FacialFeature}, texture \cite{EffectsofVisualRealism, zell2015stylize, effectsofskincolour}, lighting \cite{crowdlight, Perceptionoflighting, light3}, expressions and emotions \cite{higgins2021remotely, photorealismavoidtheuncanny, EffectofMotionTypeandEmotions}, level of detail \cite{effectofmotionandbodyshape, ImpactofVaryingResolution, photorealismavoidtheuncanny, Perceptionofvirtualcharacters, empathicresponses}, facial proportions \cite{FacialFeature, resize}, and personalization \cite{VirtualTwin, higgins2021remotely}. Fidelity assessments are typically paired with evaluations of the uncanny valley effect as further explained in the next subsection on affinity. Prior research shows that mismatches between an avatar’s appearance, animation, or voice often increase eeriness and reduce user experience quality. For an extensive review of visualization of avatars and agents in AR and VR, we refer to \cite{review_avatarvisualization}. Achieving high fidelity is related with the capabilities of the tools and methods used and the application type. For instance, Ma and Pan \cite{VisualFidelity} reported a preference for cartoonish over photorealistic avatars in VR, diverging from other studies \cite{photorealismavoidtheuncanny, IsPhotorealismImportant} that argue for photorealism. Kokkinara et al. \cite{kokkinara2015animation} suggest that virtual faces are perceived as more appealing when higher levels of animation realism are provided. Amadou et al.~\cite{NabilaAmadou} demonstrated that mismatched fidelity between appearance and animation worsens user impressions, especially in photorealistic render styles. Building on these findings, our study uses both realistic and cartoon avatars and matches them with appropriate fidelity of facial animations using performance capture based tracking. 

\subsubsection{Affinity}: Previous research has extensively explored the link between human likeness in virtual characters and the \textit{Uncanny Valley Effect} \cite{mori2012uncanny}. To quantify this effect, the concept of affinity has been used to assess how appealing, eerie, or attractive a character appears \cite{photorealismavoidtheuncanny}. Human likeness has traditionally been evaluated through facial resemblance and visual characteristics. For example, MacDorman et al. \cite{toorealforcomfort} examined how facial proportions, skin texture, and level of detail (LOD) impact perceived eeriness, human likeness, and attractiveness. The study found that stylized textures (e.g., line-drawing) paired with high LOD increased eeriness, supporting Uncanny Valley theory. Higgins et al.~\cite{photorealismavoidtheuncanny} investigated the effect of using photorealistic MetaHuman avatars \cite{metahuman}, showing that advanced realism led to avatars being rated as more attractive, human-like, and less eerie. This aligns with findings from \cite{IsPhotorealismImportant}, suggesting current technology can mitigate uncanny valley effects in virtual humans. Higgins et al. further noted that higher geometric and texture detail improved affinity ratings, underscoring the advantage of modern high-fidelity avatars. However, their study focused solely on appearance and did not examine affinity toward avatars of familiar individuals, as our study does. Zibrek et al.~\cite{role} argued that affinity depends on avatar behavior; when behavior is plausible, visual realism can enhance attractiveness. Their work also used unfamiliar characters, preventing viewers from associating avatars with real people.

\subsubsection{Social presence}: Avatars are often credited with enhancing users' \textit{sense of presence} in virtual environments, encompassing both self-presence \cite{SelfpresentationtowardsAvatar} and social presence \cite{IsPhotorealismImportant,Mal2024_oddone}. Presence refers to the psychological experience of "being there" \cite{deepvr}, described as a \textit{"perceptual illusion of non-mediation"} \cite{conceptpresence}. Kwan \cite{presence} defines it as \textit{"a psychological state in which the virtuality of experience goes unnoticed"}, where users become immersed and disregard the technological mediation. Presence is typically discussed in two dimensions: self-presence— when users cannot distinguish between their real and virtual selves and co-presence (or social presence), which refers to the perception of virtual social actors as real. Mystakidis \cite{2022metaverse} adds that social presence goes beyond avatars, encompassing the recognition of the individuals behind them and the sense of shared existence. While some studies explore the role of avatars in fostering self-presence \cite{resize, SelfpresentationtowardsAvatar}, others focus on their contribution to social presence in virtual social platforms \cite{vrchat,IsPhotorealismImportant,nicetomeetyou}. Findings vary due to differences in emphasis and experimental design. For example, Garau \cite{garau2003impact} reports that mismatches between appearance and behavior reduce presence. Pakanen et al. \cite{nicetomeetyou} found that realistic avatars can enhance both self and social presence, although the fidelity of their avatars was limited. Zibrek et al. \cite{role} similarly observed that appearance realism increases social presence when paired with realistic behavior, while a later study \cite{IsPhotorealismImportant} suggests avatar style alone does not significantly impact social presence. In contrast to these studies, which used generic or unfamiliar avatars, our work features personalized avatars of familiar individuals. We adopt the definition of social presence from \cite{2022metaverse} and apply the questionnaire from \cite{IsPhotorealismImportant}, omitting two items specific to VR setups to better suit our experiment.

\subsection{Perception of Realistic Digital Doubles}
\label{sec:perceptionofrdd}
Human digital doubles are being used prevalently by the industry in recent years. Fortnite (a popular live service video game that uses stylized cartoon looking characters) used many celebrities' digital doubles in in-game events. Big budget AAA games, like \textit{Death Stranding} starred realistic digital doubles of actors and actresses. Seymour et al. \cite{meetmike} showcased an interactive application where people could interact with the realistic looking digital double of a real-person in VR. There has been several works that investigated personalized avatars or as we conceptualize them as digital doubles, such as- \cite{VirtualTwin, Kim2023, Plausibility_wolf_mal, Seymour2022, VisualFidelity, higgins2021remotely, Mal2024_oddone} as also referred in the previous sections. However, most of these works can create stimuli with limited realism. The stimuli is often created using avatar customization tools such as Reallusion, Mixamo or Avatar SDK. Although they provide fast creation, the perceived realism is low. Some works rely on photogrammetry systems though the results are still far from being realistic.  Creating high realism/high fidelity 3D avatars as stimuli requires high-end appearance and performance capture systems and manual artistic effort. Salagean et al. \cite{VirtualTwin} use customized avatars for participants by scanning their faces with photogrammetry. They compare their digital doubles with different levels of photorealism and personalization (self and generic). However, the stimuli they created is of moderate quality, having high human resemblance but low graphical resolution as categorized in Fraser et al. \cite{FRASER2024100082}. Furthermore, unlike in our experiment, they neither incorporated high-quality facial animation in their experiment (making the digital doubles lack vivid facial expressions), nor they had varying levels of avatar familiarity. High-end systems and manual artistic effort might not always be accessible and they might be costly for such perceptual studies. Recently, there are examples of more accessible pipelines for creating realistic and high fidelity digital humans such as Unreal Engine MetaHumans with high human resemblance and high graphical resolution. A few of the previous research work used this pipeline, however that was for the perception of generic avatars not digital doubles \cite {photorealismavoidtheuncanny, EmpathyHiggins2022, NabilaAmadou}. Fraser et al. \cite{FRASER2024100082} investigated the perception of realistic digital doubles using MetaHumans but they did not study the familiarity dimension in link with cartoonish and realistic styles. Our work is the first one to explore this and we hope that it leads to more understanding on the perception of realistic digital doubles.

\section{Methodology}
We investigated the effect of digital double representations on the perception of self/other identification, perceived realism, affinity and social presence. Although these aspects were studied in previous works regarding realistic virtual humans \cite{photorealismavoidtheuncanny,NabilaAmadou,IsPhotorealismImportant}, they were not analyzed in the context of digital doubles/avatars of real people. In addition, we looked into the effect of familiarity of the digital doubles. In particular, how people perceive their own and other people's digital doubles, which was not investigated before. In this section, we elaborate on our experiment methodology.

\subsection{Study Design}
We define a $2\times3$ within-subject study design where we have two independent variables (IV) - \textbf{[IV\textsubscript{1}]}: style (with two levels- realistic and cartoon) and \textbf{[IV\textsubscript{2}]}: familiarity (with three levels- self, acquaintance and unknown). The dependent variable groups (DV) are- \textbf{[DV\textsubscript{1}]}: (self and other) identification- that tells us how well the participants identify the digital doubles as their own, acquaintance or a stranger; \textbf{[DV\textsubscript{2}]}: perceived realism- measures whether generated animation and appearance quality is well-perceived by the users; \textbf{[DV\textsubscript{3}]}: affinity- measuring the human-likeness of the avatars; \textbf{[DV\textsubscript{4}]}: social presence- that measures the sense of being together with a real person. Each DV group consists of two or more dependent variable scales (see Table \ref{tab:self_question}).

\subsection{Hypotheses}
In line with the previous works \cite{photorealismavoidtheuncanny, NabilaAmadou,IsPhotorealismImportant, FRASER2024100082} on the perception of realistic virtual humans, we expect that higher appearance realism will lead to higher level of self/other identification, perceived realism, affinity and social presence. On the other hand, we expect that digital doubles with familiar faces (i.e., self face: high familiarity, interaction partner (acquaintance)'s face: moderate familiarity and unknown person's face: no familiarity, following Campbell et al. \cite{Familiarity_CAMPBELL2020107415}) will lead to lower level of identification, perceived realism, affinity and social presence as people are more sensitive when it comes to their own digital doubles. Therefore, we investigate the following hypotheses:
\begin{itemize}
    \item[\textbf{[H1a]}] Higher appearance realism of digital doubles leads to a higher level of self/other-identification than lower appearance realism digital doubles.
    \item[\textbf{[H1b]}] Digital doubles with familiar faces lead to a lower level of self/other-identification than digital doubles with unfamiliar faces.
    \item[\textbf{[H2a]}] Higher appearance realism of digital doubles leads to higher perceived realism than lower appearance realism digital doubles.
    \item[\textbf{[H2b]}] Digital doubles with familiar faces lead to lower perceived realism than digital doubles with unfamiliar faces.
    \item[\textbf{[H3a]}] Higher appearance realism of digital doubles leads to a higher level of affinity than lower appearance realism digital doubles. 
    \item[\textbf{[H3b]}] Digital doubles with familiar faces lead to a lower level of affinity than digital doubles with unfamiliar faces.
    \item[\textbf{[H4a]}] Higher appearance realism of digital doubles leads to a higher level of social presence than lower appearance realism digital doubles.
    \item[\textbf{[H4b]}] Digital doubles with familiar faces lead to a lower level of social presence than digital doubles with unfamiliar faces.
\end{itemize}

\subsection{Participants}
This study investigates not only participants' perceptions of their own avatars but also how familiar pairs perceive each other’s avatars through natural, face-to-face interactions reflective of daily life. To this end, we recruited participants who knew each other in real life (e.g. friends, classmates, colleagues, or family) and tested them in pairs. Each participant served as the other's \textit{interaction partner} whom they were already acquainted with. We recruited 11 such pairs (22 participants total), consisting of university students and staff. Participants ranged in age from 18 to 34; twelve identified as male, eight as female, and two as non-binary. The study received ethical approval from the XX University review board, and all participants provided informed consent for data collection and the creation of their digital doubles. Permission was also obtained for use of participant images in this manuscript. Relationships within each pair were either "classmate" or "co-worker." To control for unfamiliarity in the "stranger" condition, we created two pools: pool 1 (classmates) and pool 2 (university staff). Participants within each pool knew each other but not members of the other, allowing us to assign a truly unknown person’s avatar from the opposite pool in the survey.

\subsection{Avatar Creation}
To create the two avatar style conditions for the human digital doubles, we used MetaHuman (MH) \cite{metahuman} in Unreal Engine 5 for high-fidelity, realistic avatars, and ReadyPlayerMe (RPM) \cite{readyplayerme} for stylized avatars. Calibration videos were recorded using Unreal Engine's LiveLink Face app \cite{livelinkface} in \textit{MetaHuman Animator Mode}, capturing neutral expressions from front and side views, as well as teeth exposure. Unlike the \textit{Apple ARKit Mode}, which uses 52 blendshapes, the Animator Mode captures high-quality depth data, enabling more expressive animations exclusive to MH topology. These videos were imported into Unreal Engine to create the \textit{MetaHuman}. While data import is straightforward, aligning captured facial features required manual intervention. We manually selected video frames and adjusted the face tracker to ensure accurate mesh alignment. After generating the \textit{MetaHuman}, we refined each avatar using the online MH editor, applying textures that resembled participants' faces and selecting high-fidelity hairstyles from the default library. When an exact match was unavailable, we exported similar styles to Blender for minimal edits, such as length adjustments or bun removal. On average, creating one MH avatar took approximately 2 hours, with an additional hour for animation solving and rendering.

\begin{figure}[t]
  \centering
\begin{subfigure}{0.30\linewidth}
    \includegraphics[width=\linewidth]{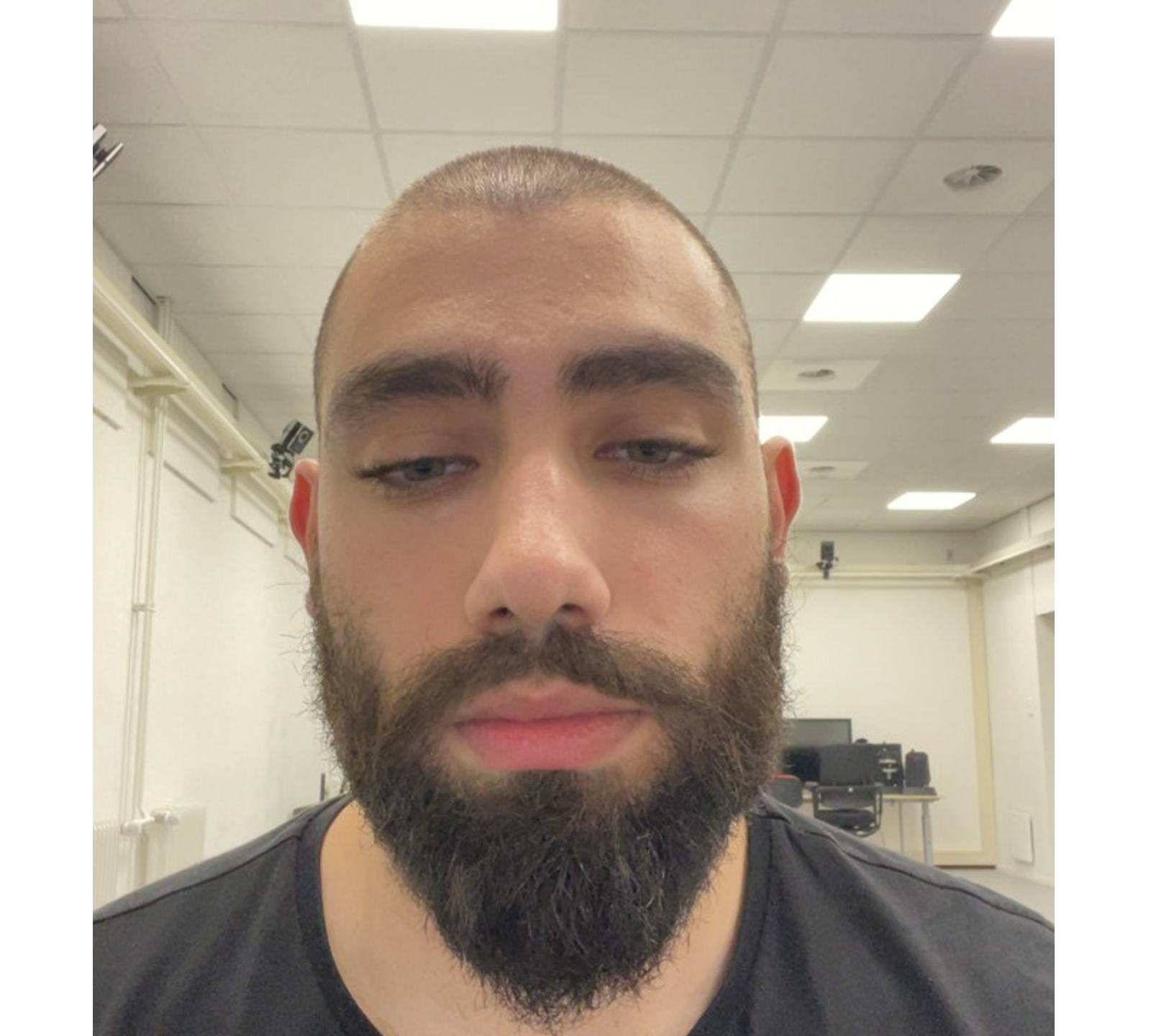}
    \caption{P1}
    \label{fig:participant1}
  \end{subfigure}
  \hfill
  \begin{subfigure}{0.30\linewidth}
    \includegraphics[width=\linewidth]{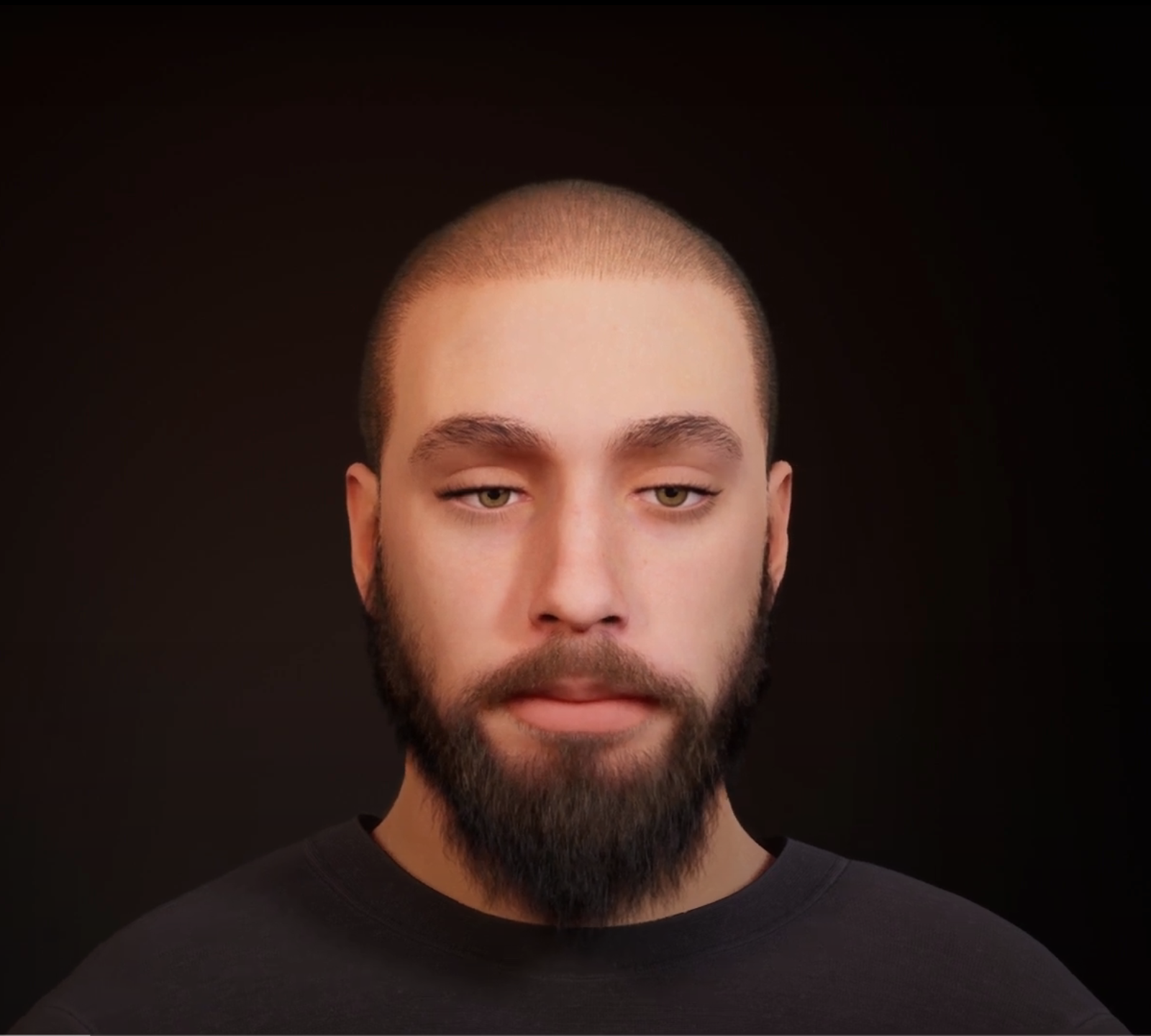}
    \caption{P1 MH Avatar}
    \label{fig:mh1}
  \end{subfigure}
  \hfill
  \begin{subfigure}{0.30\linewidth}
    \includegraphics[width=\linewidth]{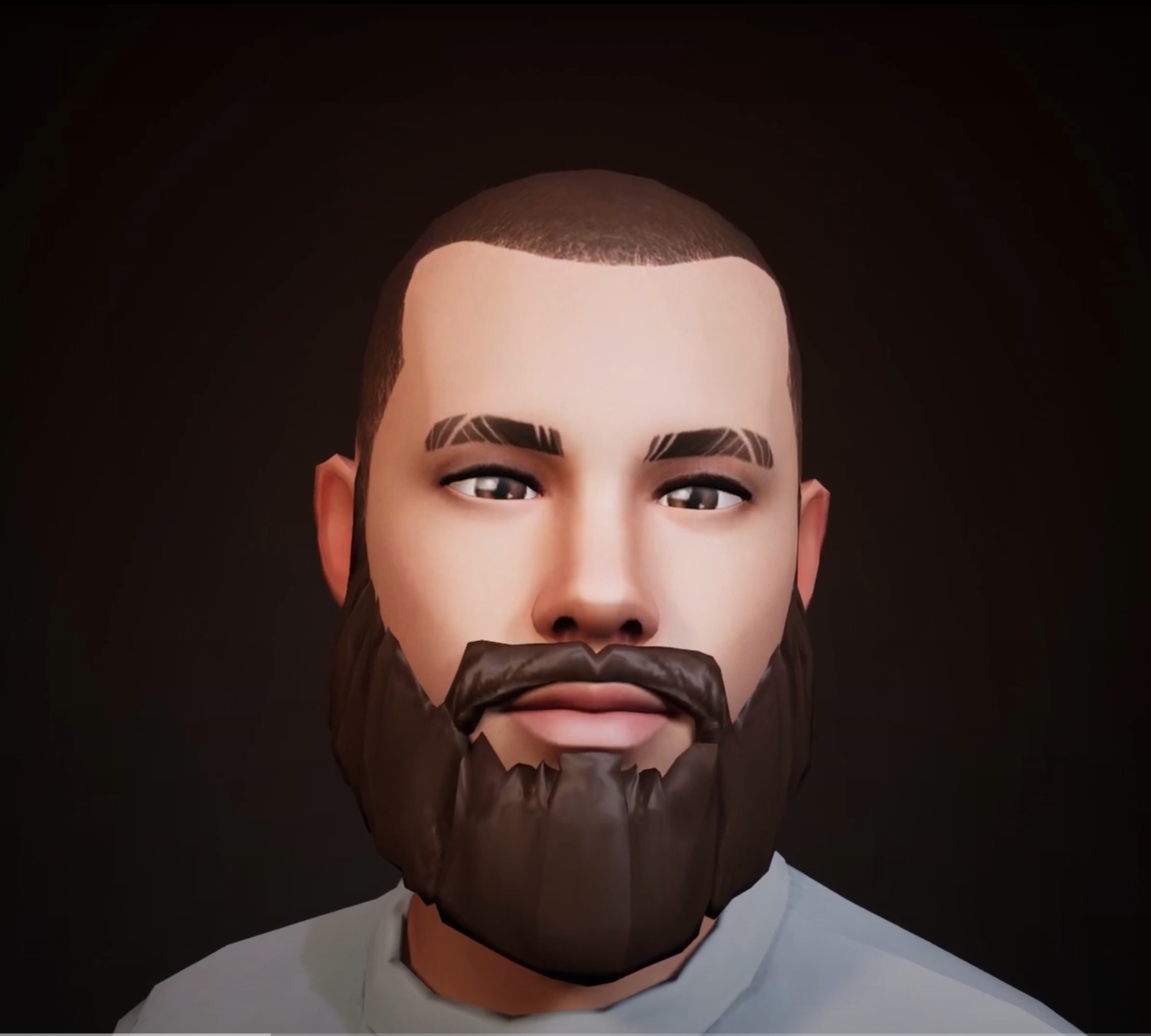}
    \caption{P1 RPM Avatar}
    \label{fig:rpm1}
  \end{subfigure}
  \hfill
  \begin{subfigure}{0.30\linewidth}
    \includegraphics[width=\linewidth]{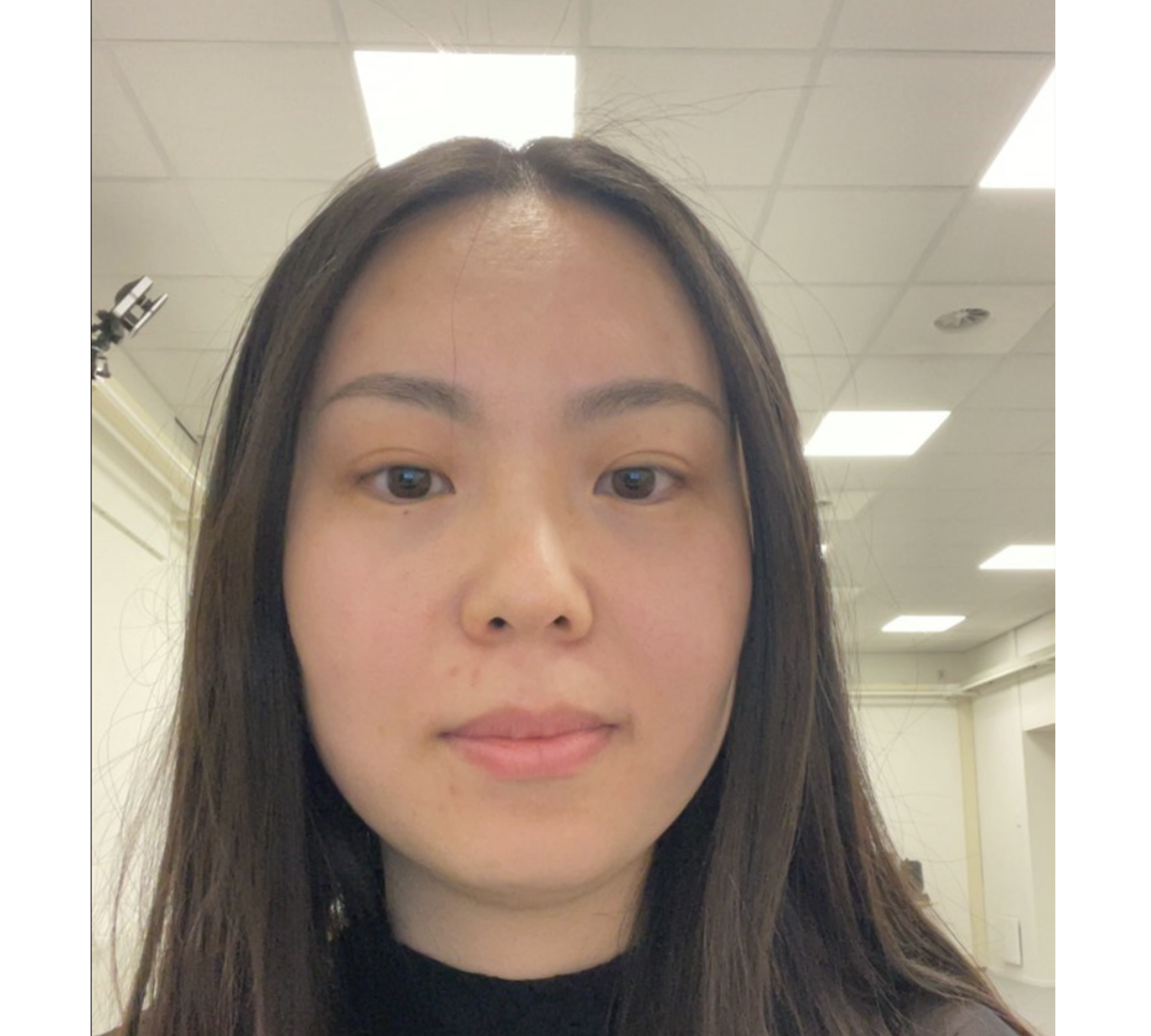}
    \caption{P2}
    \label{fig:participant2}
  \end{subfigure}
  \hfill
  \begin{subfigure}{0.30\linewidth}
    \includegraphics[width=\linewidth]{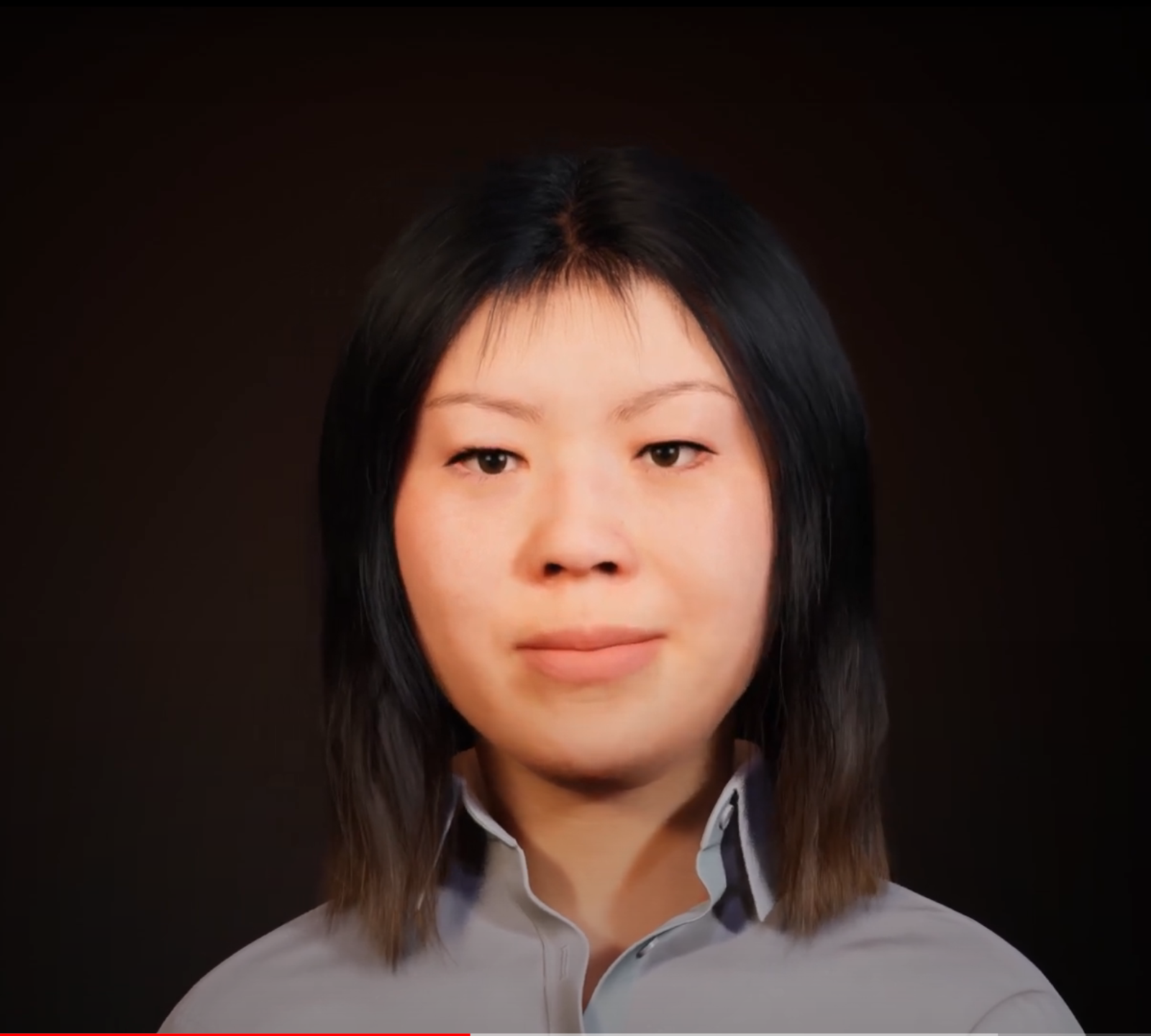}
    \caption{P2 MH Avatar}
    \label{fig:mh2}
  \end{subfigure}
  \hfill
  \begin{subfigure}{0.30\linewidth}
    \includegraphics[width=\linewidth]{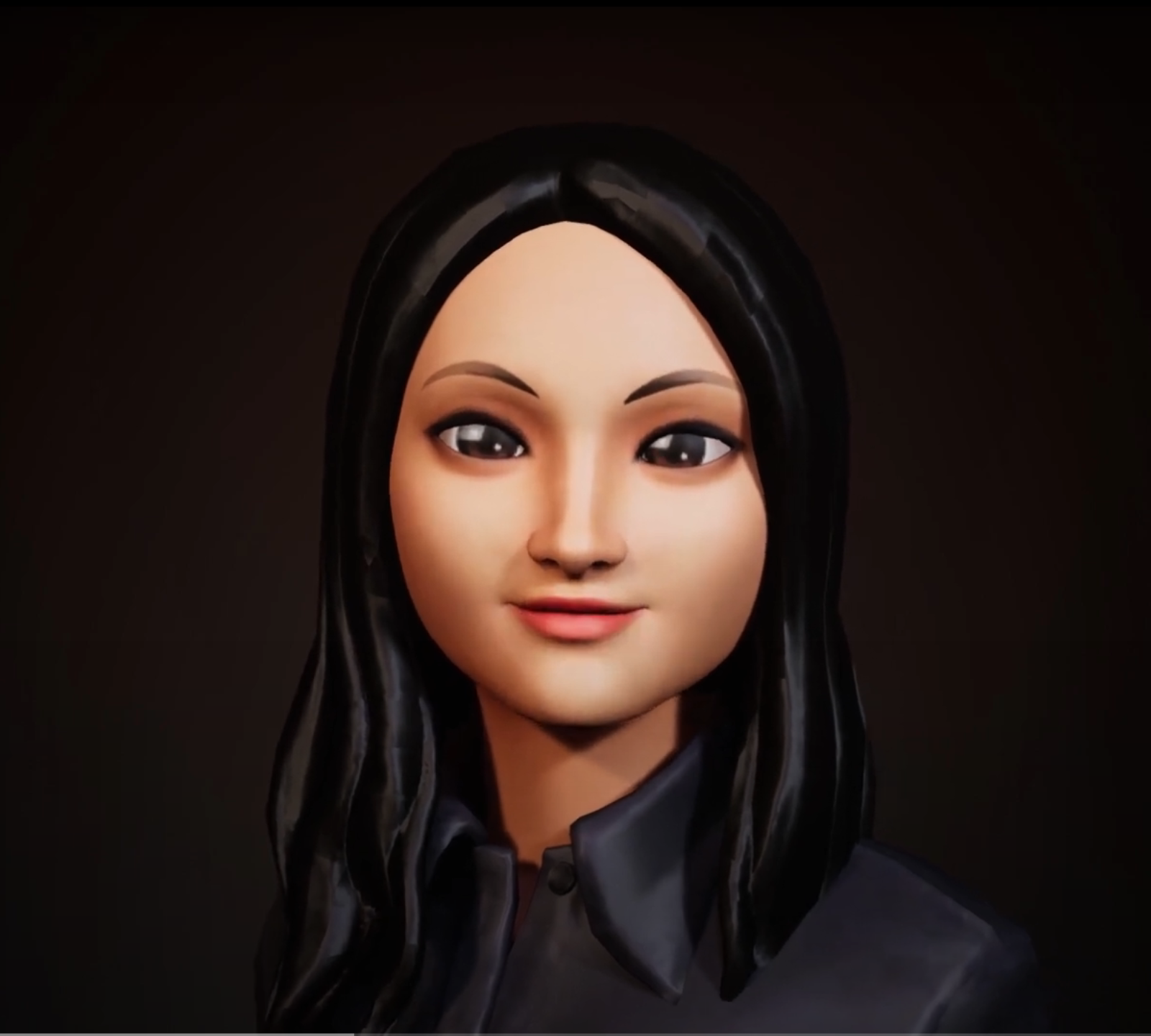}
    \caption{P2 RPM Avatar}
    \label{fig:rpm2}
  \end{subfigure}
  \caption{Examples of a pair of participants (P1 and P2) and their digital doubles created following the description presented in Sec. 3.4. Avatar Creation.}
  \label{fig:avatarcreation}
  \Description{Figure containing examples of digital doubles created for participants.}
\end{figure}

To create the stylized (i.e., cartoon) avatars, we used the free cross-game platform ReadyPlayerMe (RPM). RPM avatars are highly cartoonish and did not require calibration videos as with MetaHuman avatars. Instead, we used the RPM avatar creator to manually select features resembling participants’ faces (e.g., eye color, nose size), which were then combined to generate the final avatars. RPM avatars include blendshapes compatible with ARKit morph targets, allowing us to use the \textit{Apple ARKit Mode} for animation solving using the same performance capture videos. The creation process was straightforward; on average, each RPM avatar took about 10 minutes to build, with an additional 30 minutes for solving and rendering the animation.

The avatar talking videos used in the experiment did not include full-body renders; thus, we used default clothing assets from the MH and RPM editors without customization. As the study focused on facial appearance and animation, complex body movements were excluded. Final stimuli displayed only the head and a small portion of the upper body (see Figure~\ref{fig:avatarcreation}), minimizing the need for full-body motion capture. A default upper-body animation was applied uniformly to simulate subtle movements during breathing and speaking. Nevertheless, the questionnaire included items related to body movements and behavior to examine whether avatar style or familiarity influenced participants’ responses on identification and perceived realism.

\begin{table}[t]
\resizebox{\linewidth}{!}{
        \begin{tabular}{l|l|l}
        \toprule[2pt]
            Group & Dependent Variable & Item\\
        \midrule[2pt]
            \multirow{5}*{Identification}& Appearance & "I think that my avatar looks like me."\\
             & Talking & "I think that my avatar talks like me." \\
             & Facial Expression & "I think that my avatar’s facial expressions look like me." \\
             & Body Movement & "I think that my avatar’s body movements look like me." \\
             & Behavior & "I think that my avatar behaves like me." \\
        \midrule[1pt]
            \multirow{4}*{Perceived Realism}& Overall & "I found my avatar realistic overall."\\
             & Appearance & "I found my avatar’s appearance realistic." \\
             & Facial Expression & "I found my avatar’s facial movements realistic." \\
             & Body Movement & "I found my avatar’s body movements realistic." \\
        \midrule[1pt]
            \multirow{2}*{Affinity}& Appeal & "I found my avatar appealing, likeable."\\
             & Eerie & "I found my avatar eerie, creepy." \\
        \midrule[1pt]
            \multirow{2}*{Social Presence}& SP1 & "The thought that my avatar isn’t real crossed my mind often."\\
             & SP2 & "My avatar appears to be alive." \\
             & SP3 & "My avatar is only a computerized image, not a real person." \\
        \bottomrule[2pt]
        \end{tabular}
} 
\caption{Questions for the \textit{self-like} avatar. As for the other two avatar levels of familiarity- \textit{interaction partner (acquaintance)} and \textit{unknown person}, the statements were adapted accordingly by replacing the subject of the statement (e.g. "my interaction partner" and "existing person" respectively).}
\label{tab:self_question}
\end{table}

We consider eyeglasses an important aspect of facial identity. For participants who regularly wear them, glasses contribute significantly to their appearance and the impression they convey. Therefore, we added glasses to both the MH and RPM avatars of participants who habitually wear them. To ensure consistency, a single team member handled the avatar creation process. Example rendered stimuli are included in the supplementary video.

\subsection{Questionnaire}
\label{sec:question}
Our study measures self/other-identification, perceived realism, affinity, and social presence. Table~\ref{tab:self_question} presents the questionnaire items for self-like avatars and their associated dependent variables. Each was rated on a 7-point Likert scale (1 = strongly disagree, 7 = strongly agree). Questions for interaction partner and unknown person avatars mirrored those for self-like avatars, with wording adjusted accordingly for interaction partner and stranger. The avatar identification items were developed for this study. Items assessing perceived realism, affinity, and social presence (SP) were adapted from \cite{IsPhotorealismImportant, NabilaAmadou}, excluding VR-specific elements (e.g., place illusion, eye contact) to better suit our context.

\begin{figure*}[t]
  \centering
\begin{subfigure}{0.5\linewidth}
    \includegraphics[width=\linewidth]{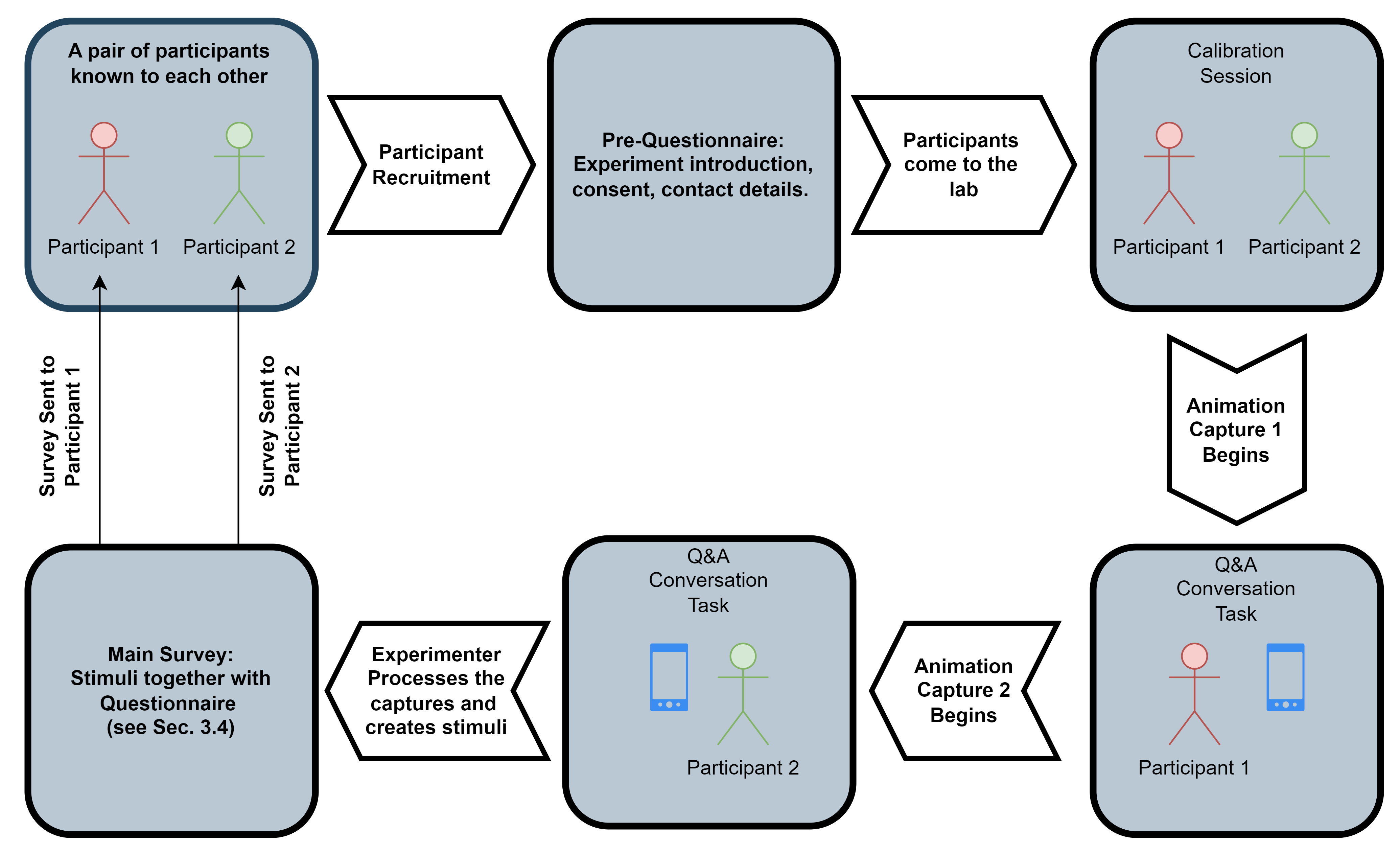}
    \caption{Experiment procedure block diagram.}
    \label{fig:ExperimentBlock}
  \end{subfigure}
  \begin{subfigure}{0.18\linewidth}
    \includegraphics[width=\linewidth,trim={1cm 0cm 1cm 0cm},clip]{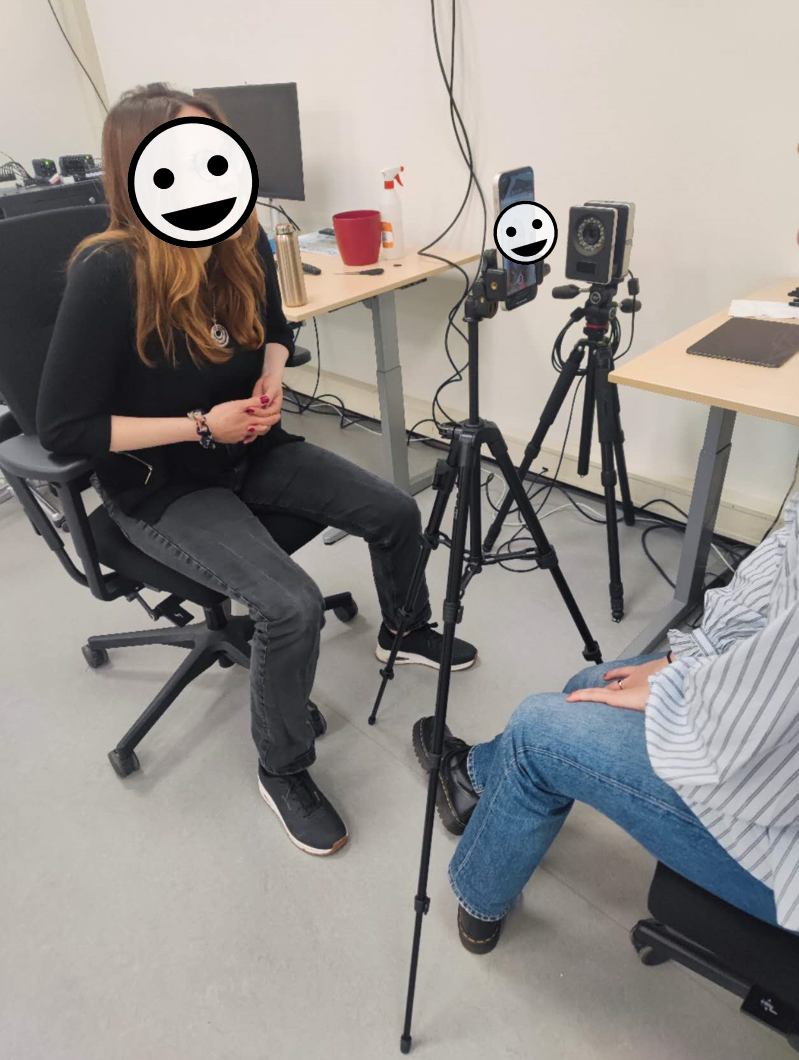}
    \caption{Q\&A conversation task setting.}
    \label{fig:experiemntreal}
  \end{subfigure}
  \caption{Figure \ref{fig:ExperimentBlock} demonstrates the step-by-step experiment procedure. In Figure \ref{fig:experiemntreal}, the two participants conducting the Q\&A conversation task. The person on the left is asking questions while the person on the right outside the frame is the respondent whose facial performance is being captured.}
  \label{fig:ExperimentProcedure}
  \Description{Figure containing details of the experiment procedure.}
\end{figure*}

\subsection{Procedure}
After the participants were recruited in pairs, they filled out a pre-questionnaire in which they learned about the experiment and provided informed consent together with their contact details for later usage. Upon arriving at the lab, each participant underwent facial capture twice. The first capture (i.e. calibration capture) was for creating the MH avatar. The second capture (animation capture) involved recording a 1-2 minute facial performance to produce the talking animation for the avatars. For the animation capture, the pair of participants engaged in a question-and-answer (Q\&A) conversation task. They sat facing each other with an iPhone 12 mounted on a tripod placed between them. One participant asked questions while the other responded, with both voices being recorded. Only the respondent's facial movements were captured by the iPhone camera. After one participant's capture was complete, they switched roles, capturing the facial animation of the other participant. As we aimed for the overall interaction to mimic the participants' real-life conversations, we allowed them to choose their conversation topics freely. In practice, the topics chosen indeed reflected typical everyday conversations, such as asking about holiday plans or deciding on a restaurant for the evening. All the captured conversations by the participants were done in a neutral emotional setting and no participant showed extreme emotional facial expressions (e.g., anger or disgust). After the stimuli were created by the experimenter, the surveys were created for respective participant pairs that included the questionnaire outlined in \ref{tab:self_question}. The questionnaire included questions regarding the two within-subject factors: avatar \textit{style} and avatar \textit{familiarity}. Each participant viewed stimuli (i.e. rendered videos) of avatars representing three levels of familiarity: self-like avatar, interaction partner (acquaintance)'s avatar, and unknown person's avatar and filled the questionnaires. The unknown person's avatar was sourced from a video of an unfamiliar participant from another pair of participants. Each familiarity level had two styles: realistic MetaHuman and stylized RPM cartoon avatar. Thus, each participant viewed a total of six avatar videos. After each video, the corresponding questionnaire was filled. Figure \ref{fig:ExperimentBlock} demonstrates the step-by-step experiment procedure.

\section{Result and Analysis}
We performed a two-way repeated measures ANOVA with avatar \textbf{style} (\textit{Realistic} MH avatar vs. \textit{Stylized cartoon} RPM avatar) and avatar \textbf{familiarity} (\textit{Self-like} avatar, \textit{Interaction Partner (acquaintance)'s} avatar, and \textit{Unknown Person's} avatar) and  as independent variables. We used the Shapiro-Wilk test to check the normality of data distribution for all measures and conditions and found non-normality in the data distribution. However, ANOVA is generally considered robust to non-normal data \cite{anova1, anova2}, and this robustness extends even to small datasets \cite{anova3}. The assumption of sphericity was checked with the Mauchly test and no violation of the assumption was found. If statistically significant differences exist, post hoc analyses were performed using the Bonferroni pairwise comparison. 

Table \ref{tab:result_i}, \ref{tab:result_r} and \ref{tab:result_u} outline all results with significant differences. Figure \ref{fig:result_i}, \ref{fig:result_r}, \ref{fig:result_u_sp} show the main effects of avatar’s style and familiarity on identification, perceived realism, affinity and social presence marked with an asterisk for statistical significance in addition to the mean values. 

\subsection{Self/Other-Identification}
\begin{figure}[t]
  \centering
\begin{subfigure}{0.30\linewidth}
    \includegraphics[width=\linewidth]{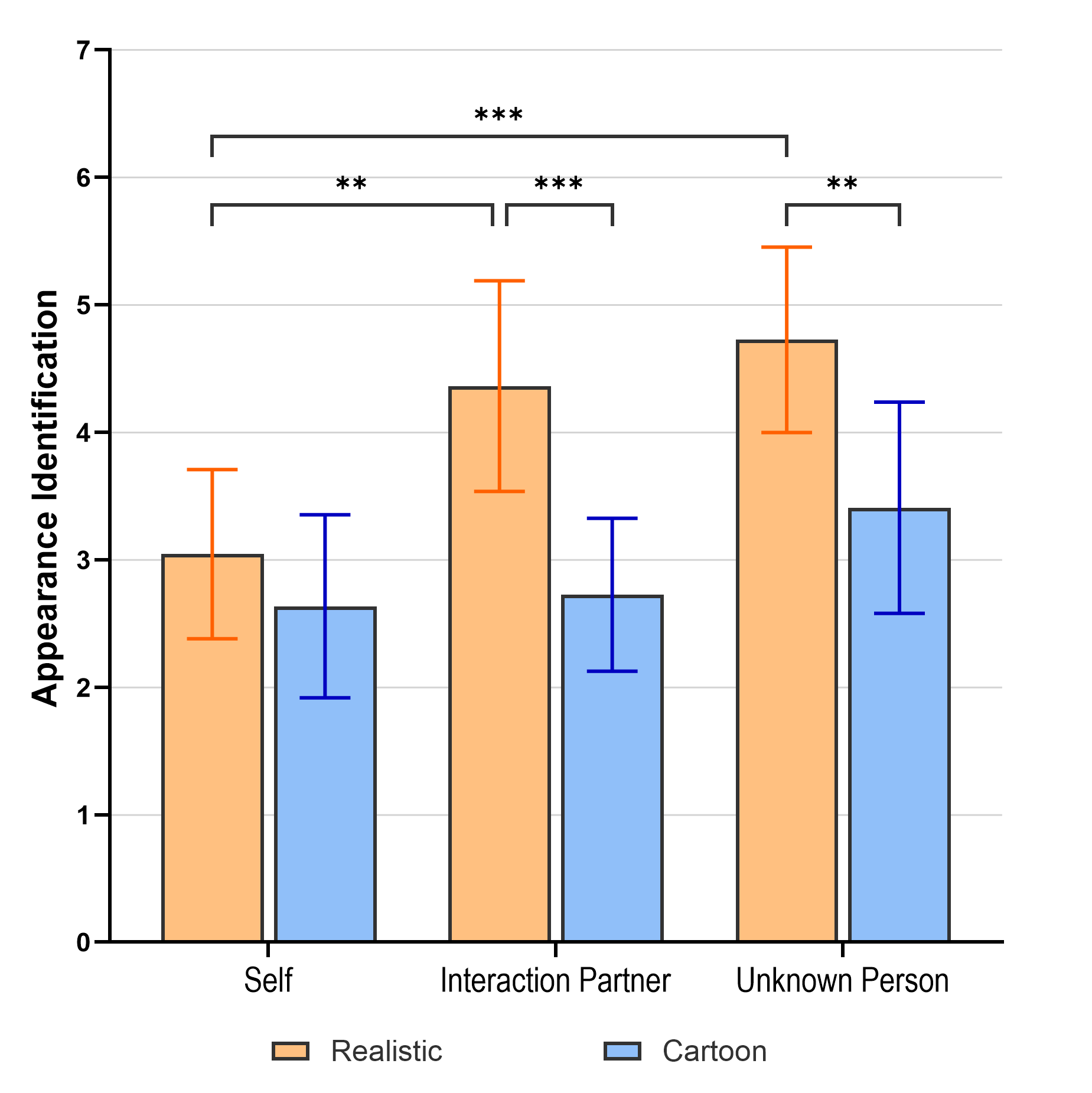}
    \caption{Appearance identification}
    \label{fig:i_appearance}
\end{subfigure}
  \hfill
\begin{subfigure}{0.30\linewidth}
    \includegraphics[width=\linewidth]{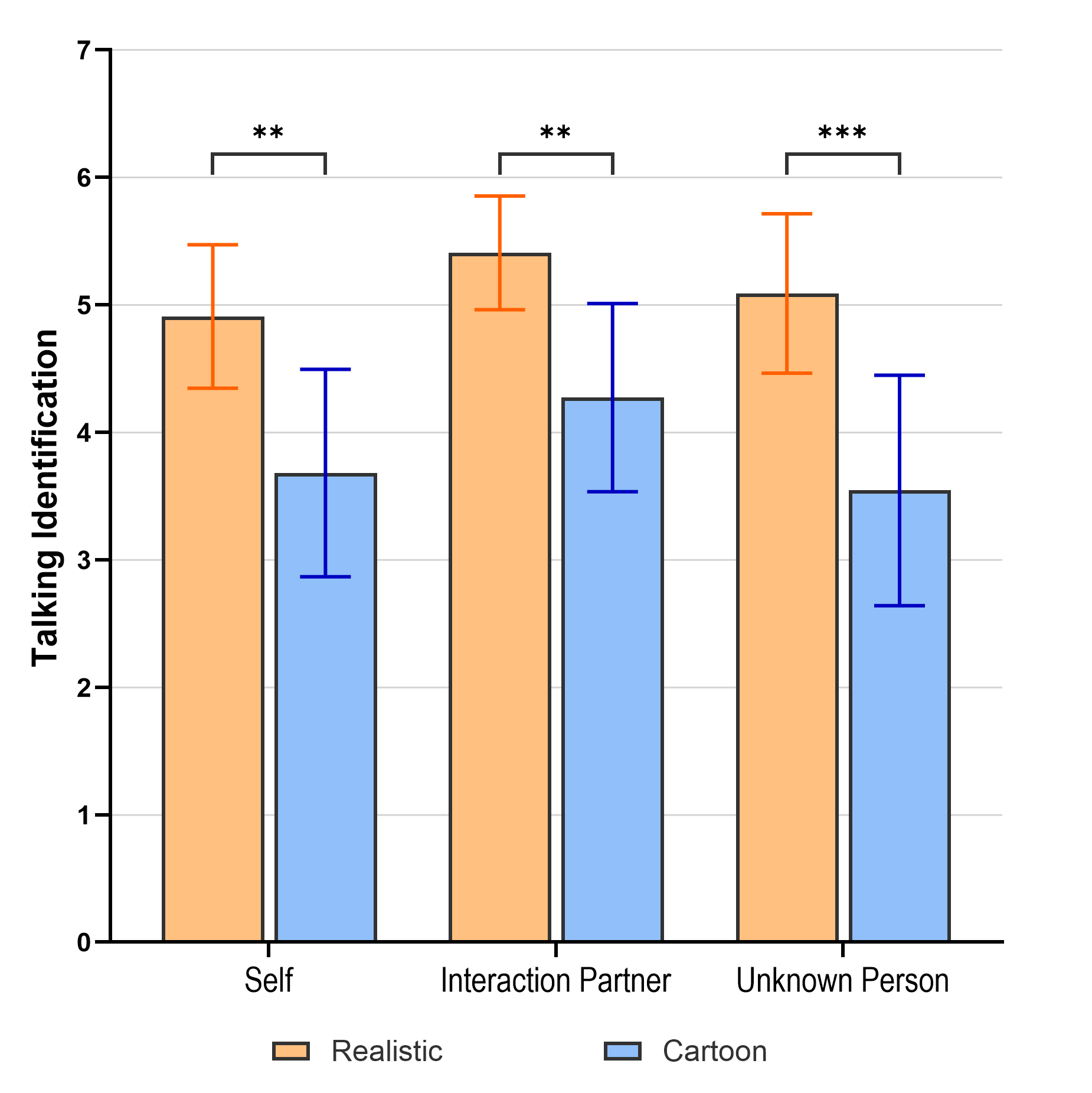}
    \caption{Talking identification}
    \label{fig:i_talking}
\end{subfigure}
  \hfill
\begin{subfigure}{0.30\linewidth}
    \includegraphics[width=\linewidth]{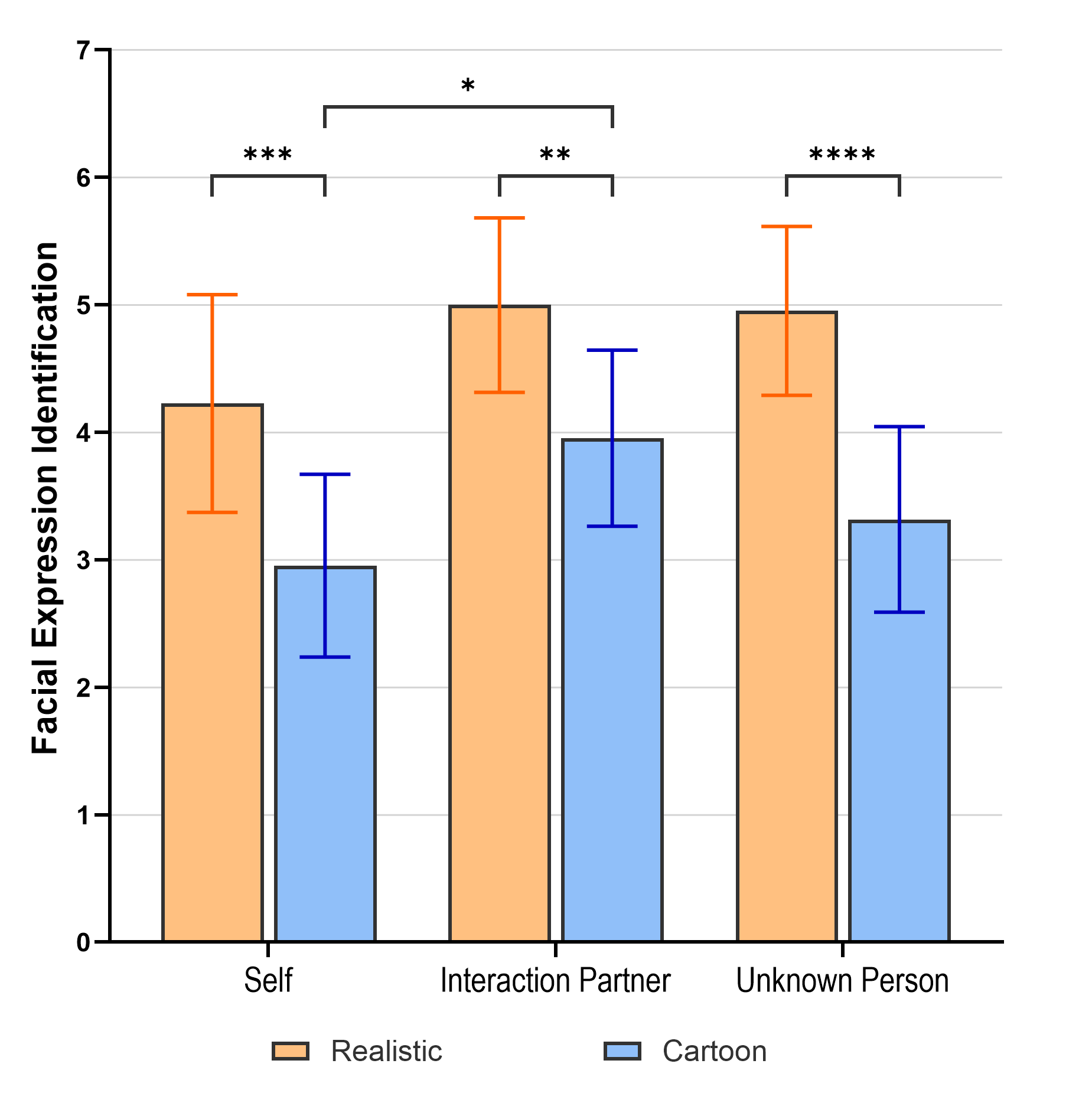}
    \caption{Facial expression identification}
    \label{fig:i_expression}
\end{subfigure}
  \hfill
\begin{subfigure}{0.30\linewidth}
    \includegraphics[width=\linewidth]{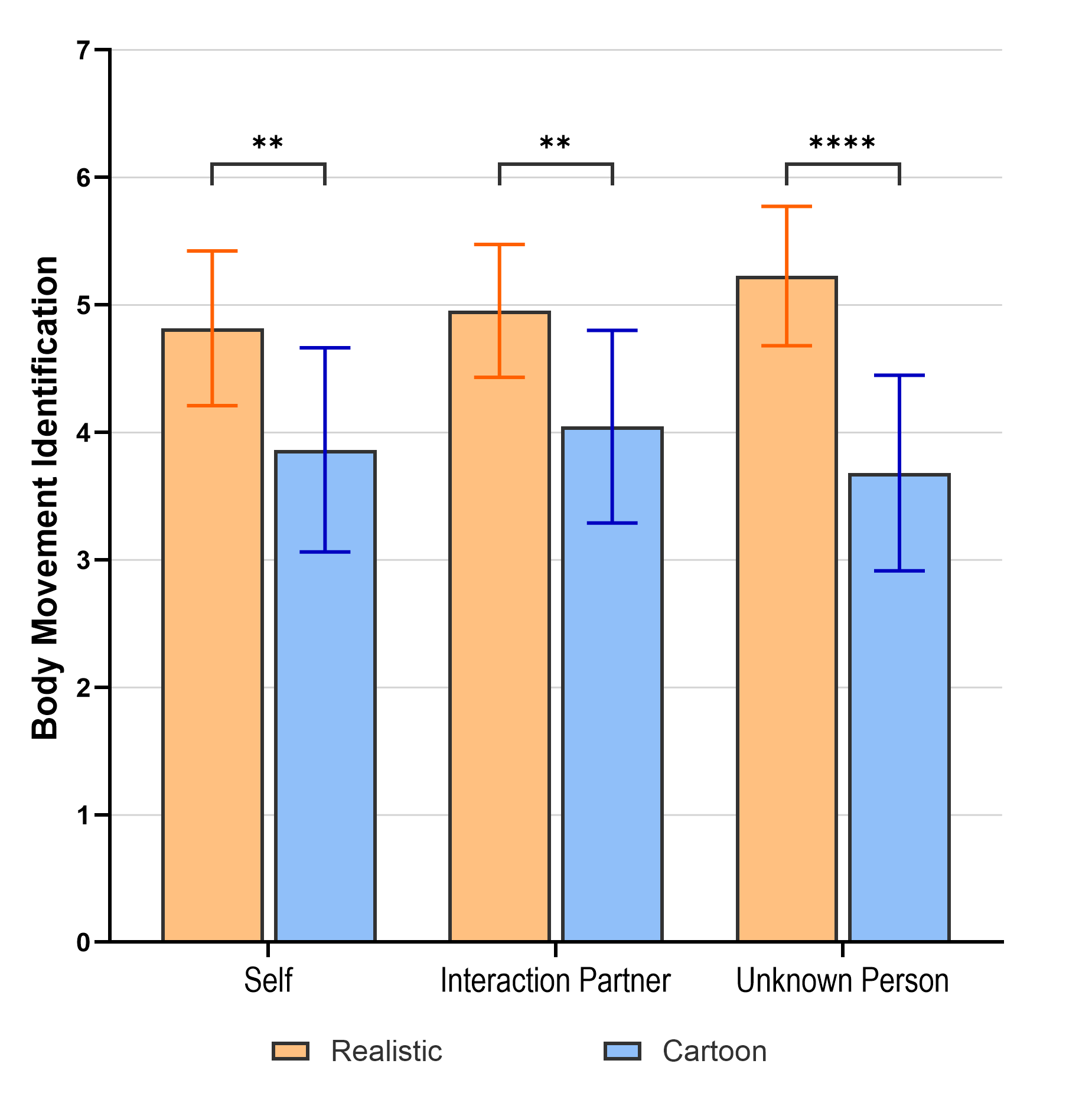}
    \caption{Body motion identification}
    \label{fig:i_movement}
\end{subfigure}
\begin{subfigure}{0.30\linewidth}
    \includegraphics[width=\linewidth]{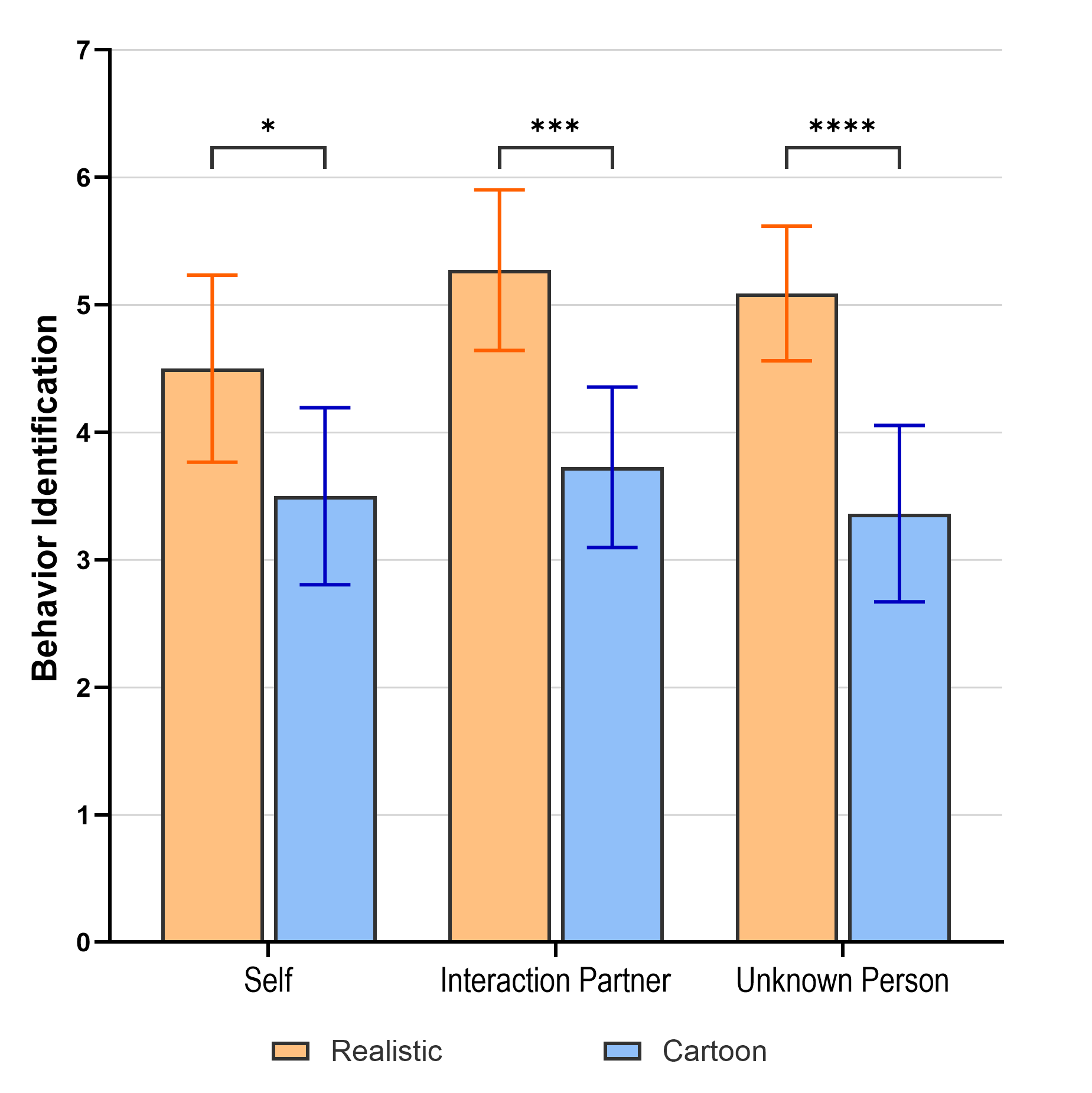}
    \caption{Behavior identification}
    \label{fig:i_behavior}
\end{subfigure}
\caption{Main effect of avatar's familiarity and style on avatar identification. In each plot of all the figures of this paper, the light orange bar corresponds to MH avatar, while the light blue bar corresponds to RPM avatar. Further, each pair of bars from left to right corresponds to self, interaction partner (acquaintance) and unknown person respectively. Lines marked with asterisks indicate the means with significant differences in Bonferroni's pairwise comparisons ($* = p<0.05$, $** = p<0.01$, $*** = p<0.001$, $**** = p<0.0001$). The error bar indicates the $95\%$ confidence interval.}
\label{fig:result_i}
\Description{Figure containing mean plots pertaining to identification.}
\end{figure}

Figure \ref{fig:result_i} presents the mean ratings for the five identification measurements, assessing how closely users perceived the digital doubles’ appearance, talking style, facial expressions, body movements, and behavior to match those of the real person. Significant main effects were observed for both avatar style and familiarity concerning identification; however, no interaction effects between these variables were found.

\subsubsection{Appearance Identification}
The two-way repeated measures ANOVA performed on appearance identification revealed a significant main effect of avatar familiarity. The appearance of the \textit{Unknown Person's} avatar is most easily identified as an existing person, while participants were more critical in recognizing familiar faces. The \textit{Interaction Partner's} avatar and the \textit{Self-like} avatar had lower identification scores in terms of appearance (\ref{fig:i_appearance}). Bonferroni pairwise comparisons showed that the \textit{Self-like} avatar was rated significantly lower than the \textit{Unknown Person's} avatar. The main effect of the avatar style was also significant. The \textit{Realistic} avatar was rated significantly higher than the \textit{Cartoon} one. The interaction effect between avatar familiarity and style on appearance identification was non-significant.

\subsubsection{Talking Identification}
A significant effect of avatar familiarity on talking identification was revealed by the two-way repeated measures ANOVA. The \textit{Interaction Partner's} avatar was rated the highest, followed by the \textit{Unknown Person's} avatar, and finally, the \textit{Self-like} avatar (\ref{fig:i_talking}). However, Bonferroni pairwise comparisons did not show any significant differences between the pairs. This discrepancy between the ANOVA and the Bonferroni pairwise results can be attributed to the differences in sensitivity between the two tests. ANOVA tests the overall effect of the independent variable across all levels, providing a more general assessment of differences. In contrast, Bonferroni pairwise comparisons are more conservative. Therefore, while the ANOVA detected a general effect of avatar familiarity on talking identification, the specific pairwise differences were not large enough to reach significance under the stringent Bonferroni correction. There was also a significant effect of avatar style on talking identification. The \textit{Realistic} avatar was rated significantly higher than the \textit{Cartoon} one. The interaction effect between avatar familiarity and style on talking identification was found to be non-significant.

\subsubsection{Facial Expression Identification}
The two-way repeated measures ANOVA indicated a significant main effect of avatar familiarity on facial expression identification. The \textit{Interaction Partner's} avatar was rated the highest, followed by the \textit{Unknown Person's} avatar, and lastly, the \textit{Self-like} avatar (\ref{fig:i_expression}). Bonferroni pairwise comparisons did not show any significant differences between the pairs. However, the main effect of avatar familiarity for the cartoon style condition revealed that the \textit{Interaction Partner's} avatar was rated significantly higher than the \textit{self-like} avatar (see Figure \ref{fig:result_i}). 
Additionally, there was a significant main effect of avatar style on facial expression identification, with the \textit{Realistic} avatar being rated significantly higher than the \textit{Cartoon} avatar. The interaction effect between avatar familiarity and style on facial expression identification was non-significant.

\begin{table}[t]
\resizebox{\linewidth}{!}{
    \begin{tabular}{l|l|l}
    \toprule[2pt]
        \multirow{2}*{\textbf{Dependent Variables}} & \textbf{Two-way repeated} & \multirow{2}*{\textbf{Post hoc (Bonferroni)}}\\
         & \textbf{measures ANOVA} & \\
    \midrule[2pt]
    \multicolumn{1}{l}{\textbf{Identification}}\\
    \midrule[1pt]
        APPEARANCE: Familiarity & $F(2, 42) = 7.910,$ & Unknown Person's Avatar is easier to be identified as an\\
         & $p = 0.001$, $\eta^2_p=0.274$& existing person than Self-like Avatar($p=0.004$).\\
        &&\\
        APPEARANCE: Style & $F(1, 21) = 11.797,$ & Realistic Avatar is easier to be identified as an existing person\\
        & $p = 0.002$, $\eta^2_p=0.360$& than Cartoon Avatar ($p = 0.002$).\\
        &&\\
        TALKING: Style& $F(1, 21) = 15.979, $ & Realistic Avatar is easier to be identified as an existing person\\
        &$p < 0.001$, $\eta^2_p=0.432$ & than Cartoon Avatar ($p < 0.001$).\\
        &&\\
        &&\\
        EXPRESSIONS: Style & $F(1, 21) = 16.165,$ &Realistic Avatar is easier to be identified as an existing person\\
        &$p < 0.001$, $\eta^2_p=0.435$ & than Cartoon Avatar ($p < 0.001$).\\
        &&\\
        MOVEMENT: Style & $F(1, 21) = 18.819,$ & Realistic Avatar is easier to be identified as an existing person\\
        & $p < 0.001$, $\eta^2_p=0.473$& than Cartoon Avatar ($p < 0.001$).\\
        &&\\
        BEHAVIOR: Style & $F(1, 21) = 31.904, $ & Realistic Avatar is easier to be identified as an existing person\\
        &$p < 0.001$, $\eta^2_p=0.603$& than Cartoon Avatar ($p < 0.001$).\\
        &&\\
    \midrule[1pt]

    \end{tabular}
}
\caption{Summary of main effects and interactions with post hoc analysis - Avatar Identification.}
\label{tab:result_i}
\end{table}
        
\subsubsection{Body Movement Identification}
No significant main effect of avatar familiarity on body movement identification was found by the two-way repeated measures ANOVA as expected (we remind that a constant body animation was used for all conditions). The \textit{Interaction Partner's} avatar was rated the highest, followed by the \textit{Unknown Person's} avatar, and lastly, the \textit{Self-like} avatar (\ref{fig:i_movement}). There was a significant effect of avatar style on movement perception. The \textit{Realistic} avatar was rated significantly higher than the \textit{Cartoon} one. The interaction effect between avatar familiarity and style on body movement identification was non-significant.
        
\subsubsection{Behavior Identification}
No significant effect of the avatar familiarity on behavior identification was observed in the results. The \textit{Interaction Partner's} avatar was rated the highest, followed by the \textit{Unknown Person's} avatar, and lastly, the \textit{Self-like} avatar (\ref{fig:i_behavior}). There was a significant effect of avatar style on behavior identification. The \textit{Realistic} avatar was rated significantly higher than the \textit{Cartoon} one. The interaction effect between avatar familiarity and style on behavior identification was non-significant.

\subsection{Perceived Realism}
Figure \ref{fig:result_r} displays the mean ratings across four scales for perceived realism: appearance, facial expression, body movement, and overall realism. There was no statistically significant two-way interaction effect between avatar familiarity and style for any of these scales. 

\subsubsection{Appearance Realism} 
\begin{figure}[t]
  \centering
\begin{subfigure}{0.35\linewidth}
    \includegraphics[width=\linewidth]{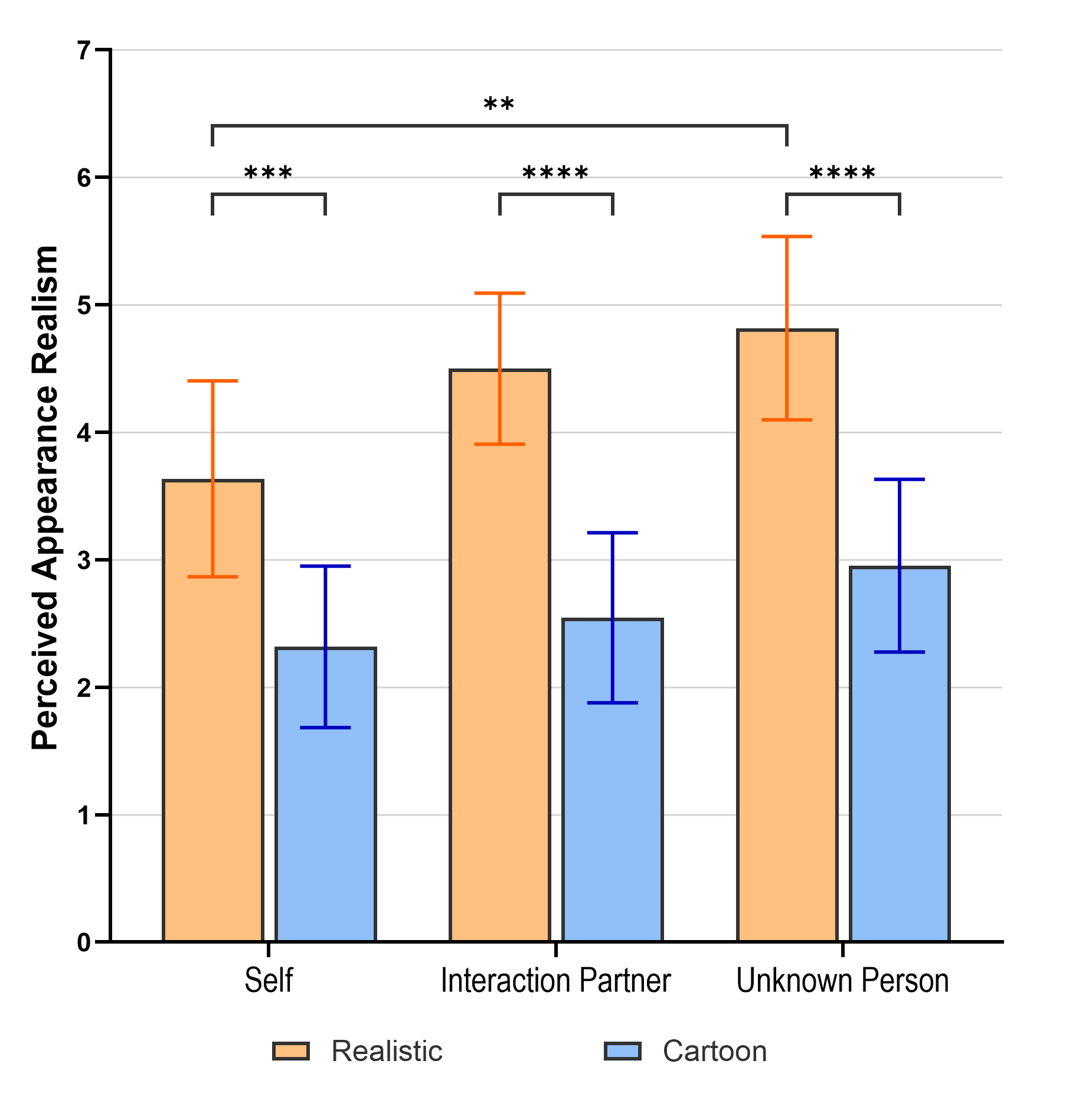}
    \caption{Appearance realism}
    \label{fig:r_appearance}
\end{subfigure}
\begin{subfigure}{0.35\linewidth}
    \includegraphics[width=\linewidth]{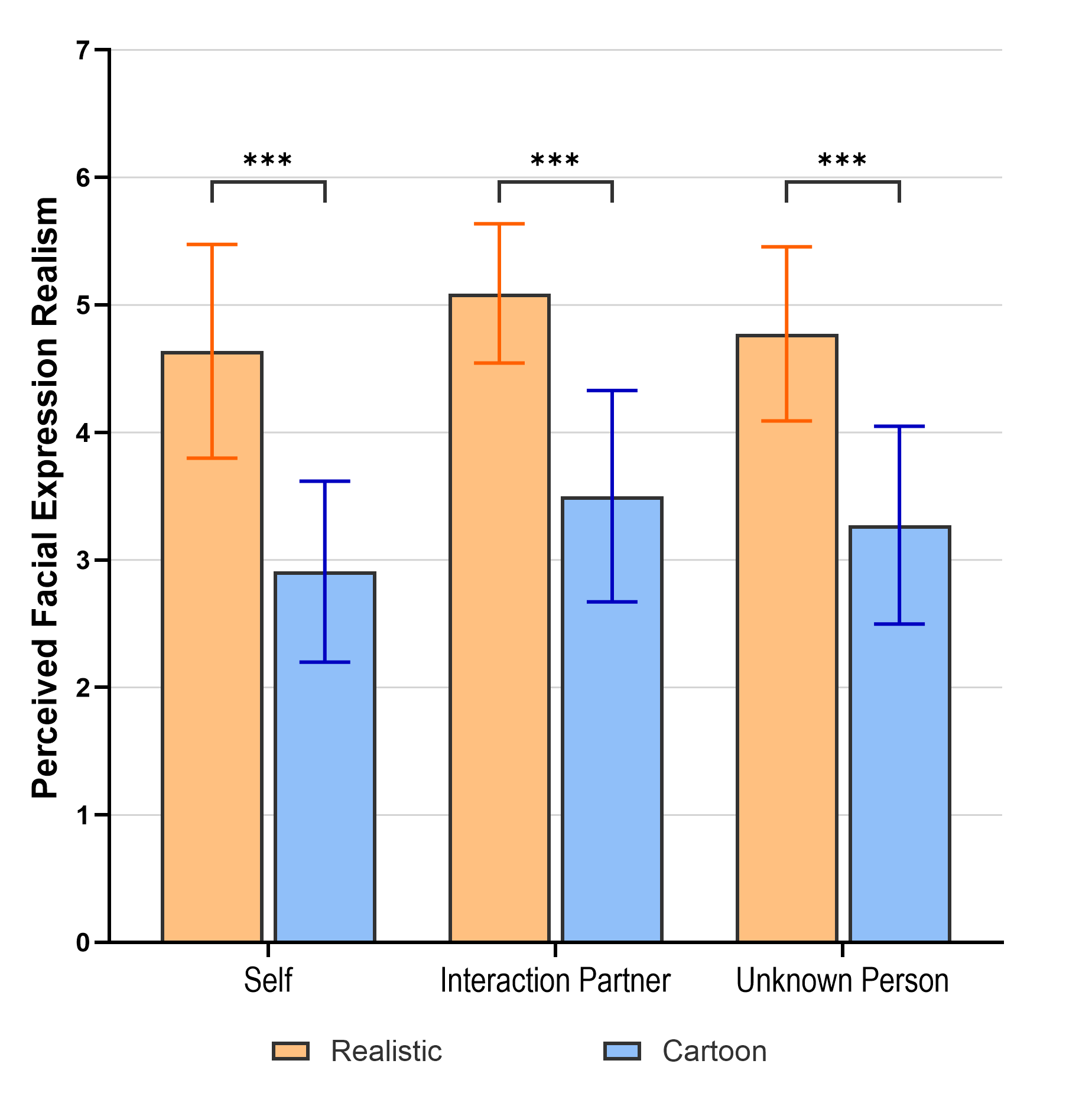}
    \caption{Expression realism}
    \label{fig:r_expression}
\end{subfigure}
\begin{subfigure}{0.35\linewidth}
    \includegraphics[width=\linewidth]{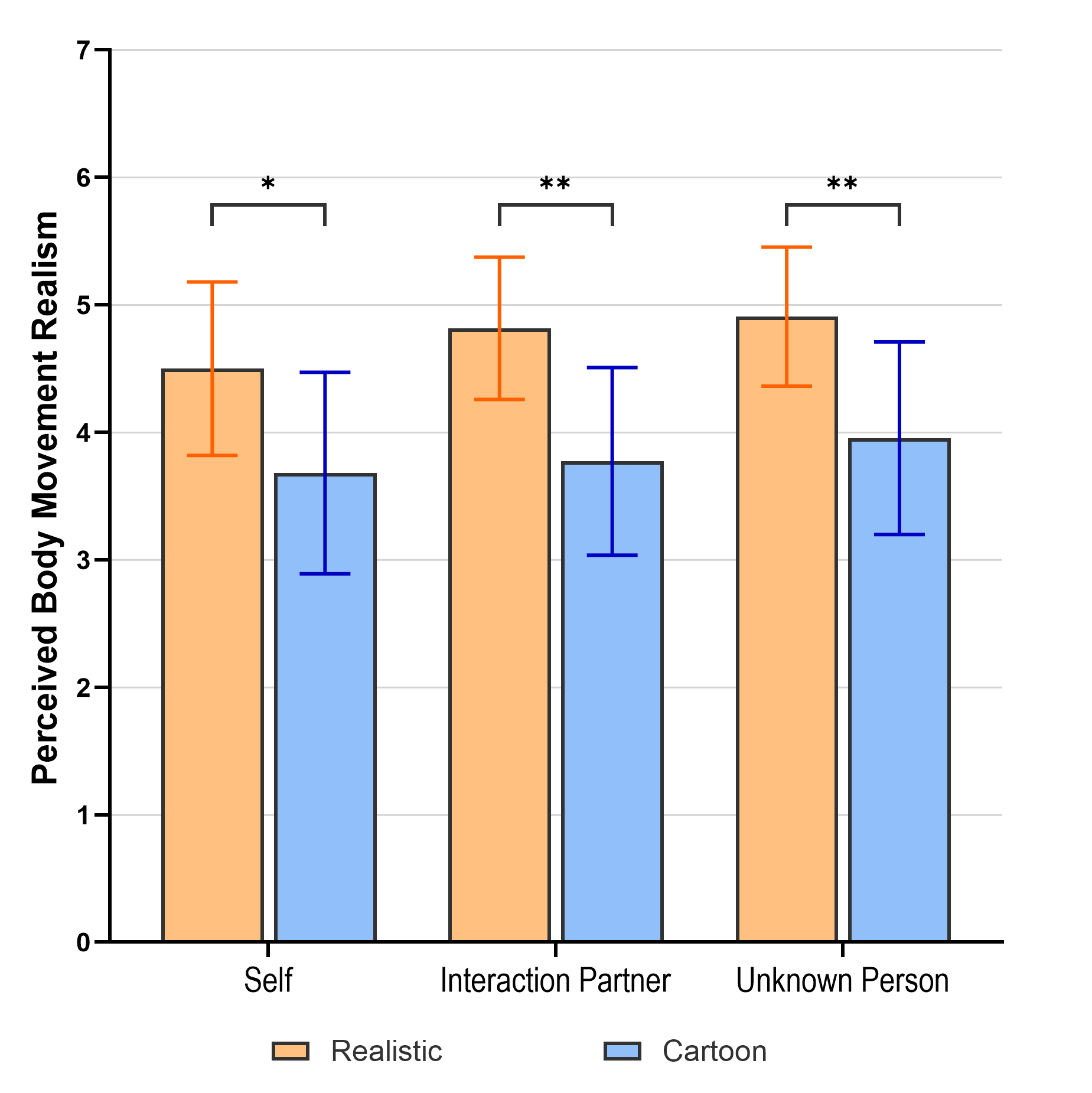}
    \caption{Body motion realism}
    \label{fig:r_movement}
\end{subfigure}
\begin{subfigure}{0.35\linewidth}
    \includegraphics[width=\linewidth]{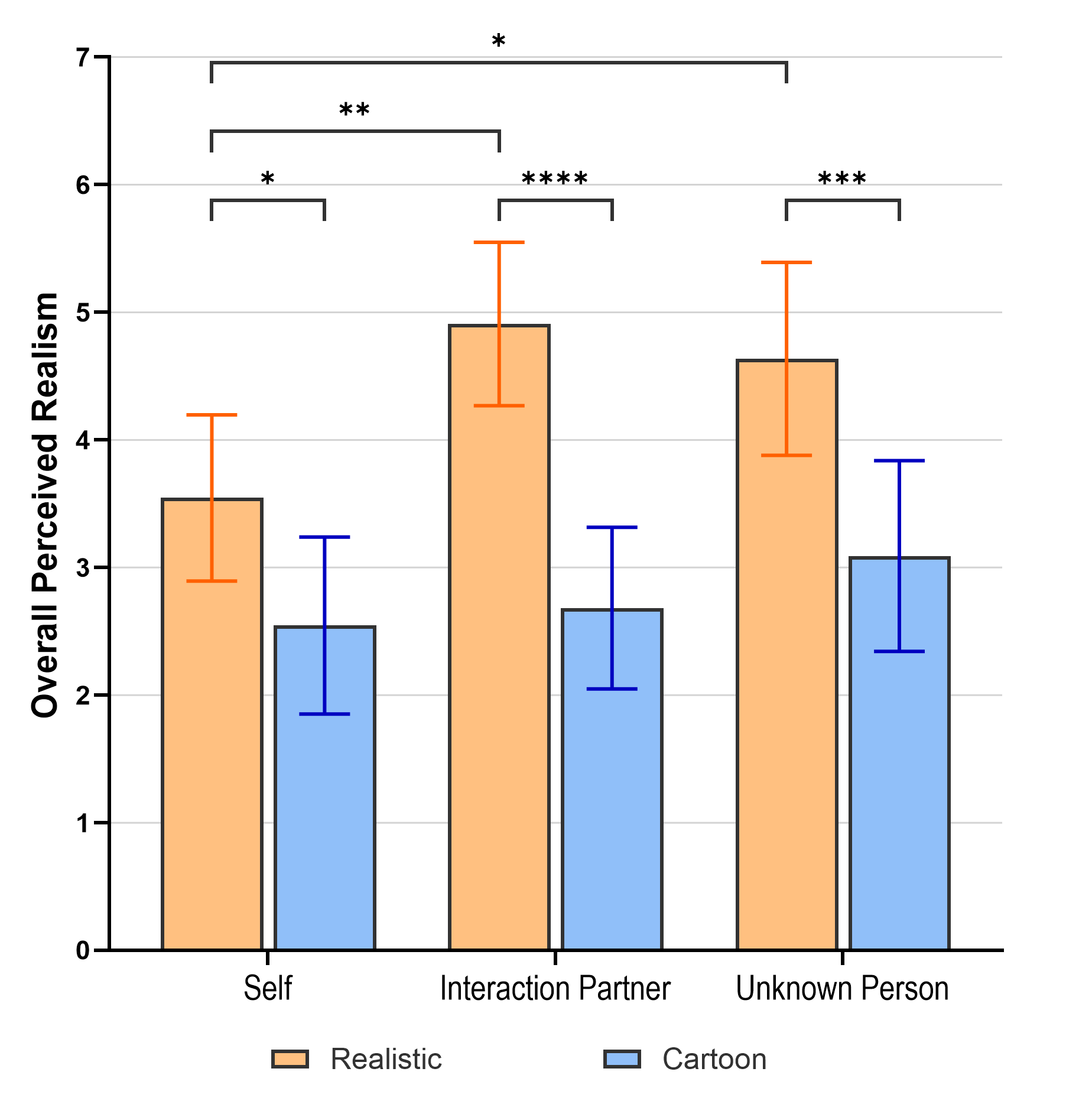}
    \caption{Overall realism}
    \label{fig:r_overall}
\end{subfigure}
\caption{Main effect of avatar's familiarity and style on perceived realism. Lines marked with an asterisk indicate the means with significant differences in Bonferroni's pairwise comparisons ($* = p<0.05$, $** = p<0.01$, $*** = p<0.001$, $**** = p<0.0001$).The error bar indicates the $95\%$ confidence interval}
\label{fig:result_r}
\Description{Figure containing mean plots pertaining to realism.}
\end{figure}

The two-way repeated measures ANOVA performed on appearance realism revealed a significant main effect of avatar familiarity. The \textit{Unknown Person's} avatar had the highest mean of appearance realism rating, followed by the \textit{Interaction Partner's} avatar, and last by the \textit{Self-like} avatar (\ref{fig:r_appearance}). Bonferroni pairwise comparisons showed that the \textit{Unknown Person's} avatar was rated significantly higher than the \textit{Self-like} avatar. The main effect of the avatar style on perceived appearance realism was also significant. The \textit{Realistic} avatar was rated significantly higher than the \textit{Cartoon} one. The interaction effect between avatar familiarity and style on appearance realism was not significant.

\subsubsection{Facial Expression Realism}
The main effect of avatar familiarity on perceived facial expression realism was non-significant.
The \textit{Interaction Partner's} avatar had the highest mean of facial expression realism rating, followed by the \textit{Unknown Person's} avatar, and lowest by the \textit{Self-like} avatar (\ref{fig:r_expression}). The main effect of the avatar style was significant. The \textit{Realistic} avatar was rated significantly higher than the \textit{Cartoon} avatar. The interaction effect between avatar familiarity and style on facial expression realism was found to be not statistically significant.

\subsubsection{Body Movement Realism}
Similar to the results of facial expression realism, body movement realism was only significantly affected by avatar style. The \textit{Realistic} avatars are rated significantly higher than the \textit{Cartoon} avatars (\ref{fig:r_movement}). However, avatar familiarity showed a non-significant effect on body movement realism. The \textit{Unknown Person's} avatar had the highest mean of body movement realism rating, followed by the \textit{Interaction Partner's} avatar, and last by the \textit{Self-like} avatar. The interaction effect between avatar familiarity and style on body movement realism was non-significant.

\begin{table}[t]
\resizebox{\linewidth}{!}{
    \begin{tabular}{l|l|l}
    \toprule[2pt]
        \multirow{2}*{\textbf{Dependent Variables}} & \textbf{Two-way repeated} & \multirow{2}*{\textbf{Post hoc (Bonferroni)}}\\
         & \textbf{measures ANOVA} & \\
    \midrule[2pt]
    \multicolumn{1}{l}{\textbf{Perceived Realism}}\\
    \midrule[1pt]
        OVERALL: Familiarity & $F(2, 42) = 8.304,$ & Unknown Person's Avatar is perceived as more realistic overall \\
         &$p < 0.001$, $\eta^2_p=0.283$&than Self-like Avatar($p = 0.002$). Interaction Partner's Avatar is\\
         & & perceived more realistic overall than Self-like Avatar($p = 0.009$).\\
        &&\\ 
        OVERALL: Style & $F(1, 21) = 30.605, $& Realistic Avatar is perceived as more realistic overall than\\
        &$p < 0.001$, $\eta^2_p=0.593$&Cartoon Avatar ($p < 0.001$).\\

        &&\\
        APPEARANCE: Familiarity & $F(2, 42) = 6.387, $ &Unknown Person's Avatar's appearance is perceived as more \\
        &$p = 0.004$, $\eta^2_p=0.233$&realistic than Self-like Avatar($p = 0.015$).\\

        &&\\
        APPEARANCE: Style & $F(1, 21) = 31.528, $ & Realistic Avatar's appearance is perceived as more realistic\\
        &$p < 0.001$, $\eta^2_p=0.600$&than Cartoon Avatar ($p < 0.001$).\\

        &&\\
        EXPRESSIONS: Style & $F(1, 21) = 20.264, $ & Realistic Avatar's facial expression is perceived as more\\
        &$p < 0.001$, $\eta^2_p=0.491$&realistic than Cartoon Avatar ($p < 0.001$).\\

        &&\\
        MOVEMENT: Style & $F(1, 21) = 16.056, $ & Realistic Avatar's body movement is perceived as more realistic\\
        &$p < 0.001$, $\eta^2_p=0.445$&than Cartoon Avatar ($p < 0.001$).\\
    \midrule[1pt]

    \end{tabular}
}
\caption{Summary of main effects and interactions with post hoc analysis - Perceived Realism.}
\label{tab:result_r}
\end{table}

\subsubsection{Overall Realism} 
There was a significant main effect of avatar familiarity found on the overall perceived realism. The \textit{Unknown Person's} avatar had the highest mean of overall realism rating, followed by the \textit{Interaction Partner's} avatar, and last by the \textit{Self-like} avatar (\ref{fig:r_overall}). The Bonferroni pairwise comparisons showed that both the \textit{Unknown Person's} avatar and the \textit{Interaction Partner's} avatar got significantly higher ratings than the \textit{Self-like} avatar. The main effect of the avatar style was also significant. The \textit{Realistic} avatar was rated significantly higher than the \textit{Cartoon} one. The interaction effect between avatar familiarity and style on overall Realism was non-significant.

\subsection{Affinity}
Figure \ref{fig:u_appeal} and \ref{fig:u_eerie} present the mean ratings for the two affinity scales: appeal and eeriness, respectively. No statistically significant two-way interaction effect was found between avatar familiarity and style for either scale.

\subsubsection{Appeal}
The two-way repeated measures ANOVA performed on appeal revealed a non-significant main effect of avatar familiarity. The \textit{Unknown Person's} avatar had the highest mean of appeal rating, followed by the \textit{Interaction Partner's} avatar, and last by the \textit{Self-like} avatar (\ref{fig:u_appeal}). The main effect of the avatar style was significant. The \textit{Cartoon} avatars were perceived as significantly more appealing than the \textit{Realistic} avatars. The interaction effect between avatar familiarity and style on appeal was not statistically significant.

\subsubsection{Eerie}
The main effect of avatar familiarity showed a statistically significant difference in perceived eeriness. Post hoc comparisons indicated that the \textit{Unknown Person's} avatar was rated significantly less eerie than the \textit{Self-like} avatar. The \textit{Interaction Partner's} avatar was rated as the second most eerie. The difference between the \textit{Interaction Partner's} avatar and the \textit{Unknown Person's} avatar was not significant. The main effect of the avatar style was also significant. The \textit{Realistic} avatars were significantly perceived as more eerie than the \textit{Cartoon} avatars. The interaction effect between avatar familiarity and style on perceived eeriness was non-significant.

\subsection{Social Presence}
\begin{figure}[t]
  \centering
\begin{subfigure}{0.32\linewidth}
    \includegraphics[width=\linewidth]{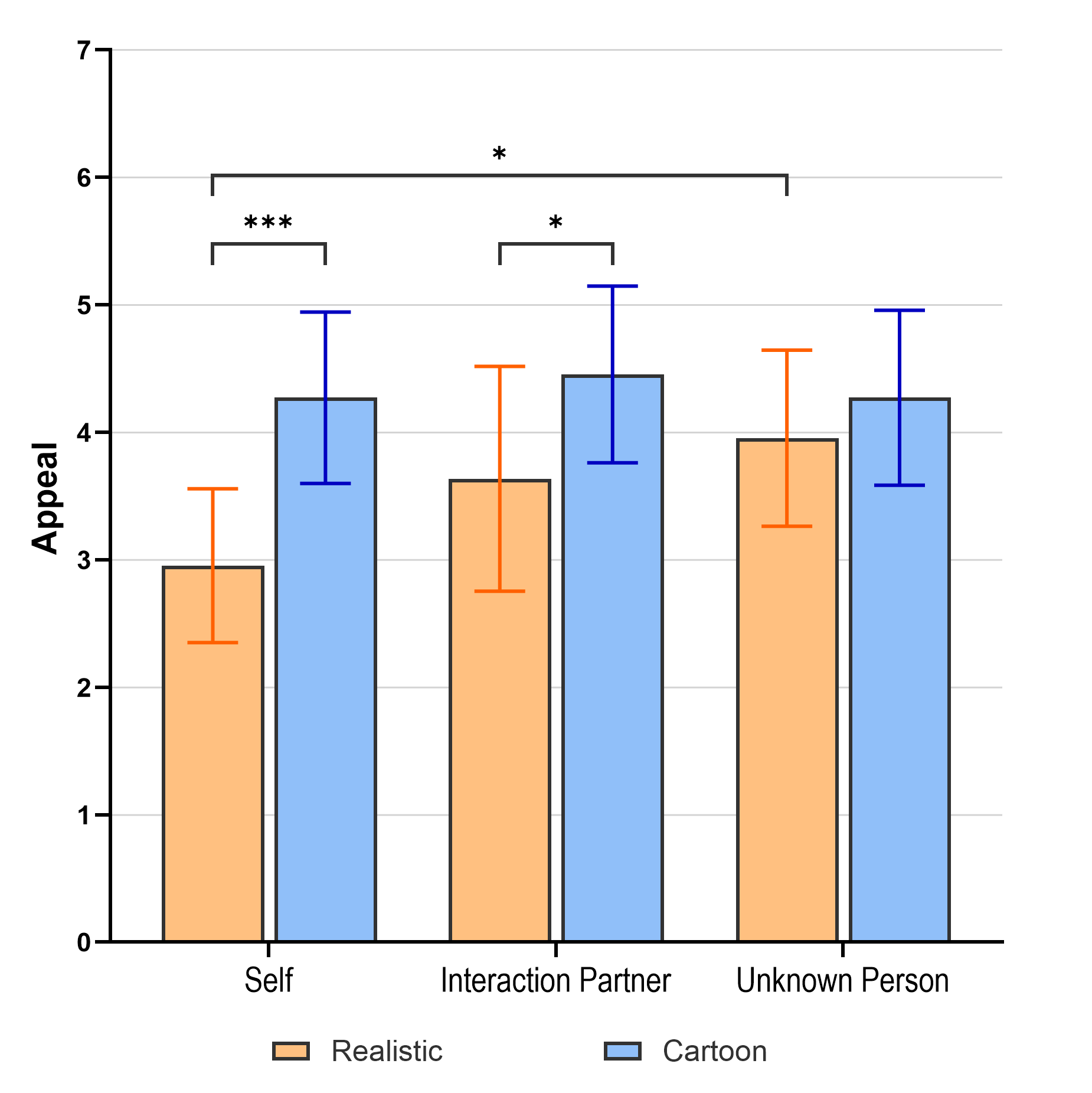}
    \caption{Appeal}
    \label{fig:u_appeal}
\end{subfigure}
  \hfill
\begin{subfigure}{0.32\linewidth}
    \includegraphics[width=\linewidth]{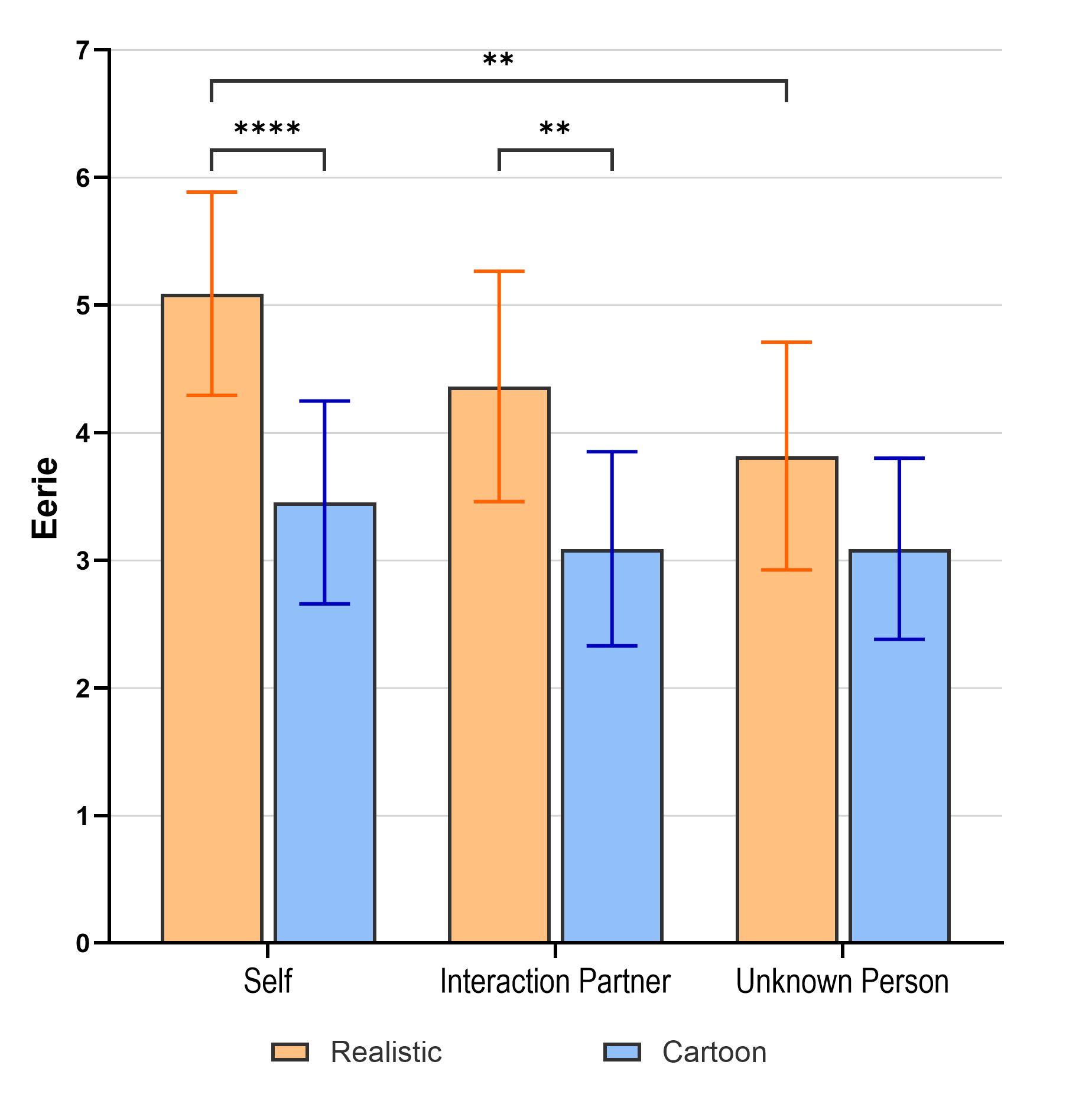}
    \caption{Eeriness}
    \label{fig:u_eerie}
\end{subfigure}
  \hfill
\begin{subfigure}{0.32\linewidth}
    \includegraphics[width=\linewidth]{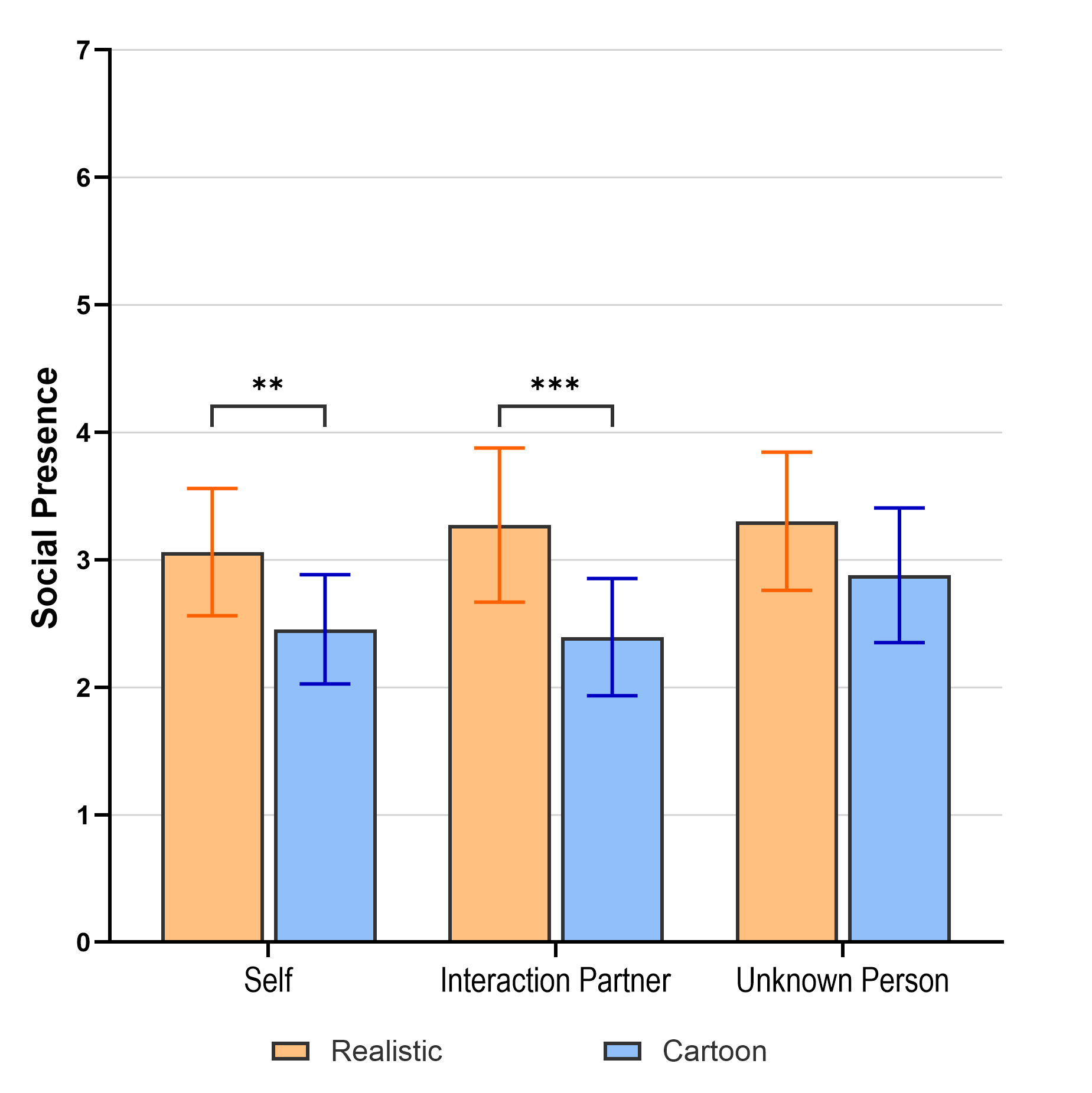}
    \caption{Social presence}
    \label{fig:sp}
\end{subfigure}
\caption{Main effect of avatar's familiarity and style on affinity (i.e. Figure \ref{fig:u_appeal} and \ref{fig:u_eerie}) and social presence (i.e. Fig \ref{fig:sp}). Lines marked with an asterisk indicate the means with significant differences in Bonferroni's pairwise comparisons ($* = p<0.05$, $** = p<0.01$, $*** = p<0.001$, $**** = p<0.0001$). The error bar indicates the $95\%$ confidence interval.}
\label{fig:result_u_sp}
\Description{Figure containing mean plots pertaining to uncanny and social presence.}
\end{figure}

The questions for the social presence scales refer to the study by Zibrek et al. \cite{IsPhotorealismImportant}. We also refer to their handling of the data for this variable by using cumulative scores of SP questions. However, since we removed two SP questions from the original scales pertaining to VR scenario that did not apply to our experiment, we tested the reliability of the remaining three questions (Cronbach's alpha: $\alpha = 0.621$), which was lower than the original five-question scale (Cronbach's alpha: $\alpha = 0.81$). Conducting the two-way repeated measures ANOVA on the cumulative scores of the three SP questions, we found that avatar familiarity had no significant effect on ratings of social presence. The \textit{Unknown Person's} avatar had the highest mean of social presence rating, followed by the \textit{Interaction Partner's} avatar, and last by the \textit{Self-like} avatar (\ref{fig:sp}). The main effect of the avatar style was significant. The \textit{Realistic} avatar was rated significantly higher than the \textit{Cartoon} avatar.

\begin{table}[t]
\resizebox{\linewidth}{!}{
    \begin{tabular}{l|l|l}
    \toprule[2pt]
        \multirow{2}*{\textbf{Dependent Variables}} & \textbf{Two-way repeated} & \multirow{2}*{\textbf{Post hoc (Bonferroni)}}\\
         & \textbf{measures ANOVA} & \\
    \midrule[2pt]
    \multicolumn{1}{l}{\textbf{Affinity}}\\
    \midrule[1pt]
        APPEAL: Style &$F(1, 21) = 5.980, $ & Realistic Avatar is less appealing than Cartoon Avatar ($p = 0.023$).\\
        &$p = 0.023$, $\eta^2_p=0.222$& \\
        & & \\
        EERIE: Familiarity & $F(2, 42) = 4.235, $ & Self-like Avatar is the most creepy one, followed by Interaction \\
        &$p = 0.021$, $\eta^2_p=0.168$&Partner's Avatar, and last by Unknown Person's Avatar. Self-like\\
        &&Avatar is rated significantly more eerie than Unknown Person's \\
        &&Avatar($p=0.032$).\\
        
        & & \\
        EERIE: Style & $F(1, 21) = 11.679,$ & Realistic Avatar is more creepy than Cartoon Avatar ($p = 0.003$).\\
        &$p = 0.003$, $\eta^2_p=0.357$& \\
    \midrule[1pt]
    \multicolumn{1}{l}{\textbf{Social Presence}}\\
    \midrule[1pt]
        \multirow{2}*{SP: Style} & $F(1, 21) = 10.115,$& Realistic Avatar is perceived with higher social presence than  \\
         &$ p = 0.005$, $\eta^2_p=0.325$& Cartoon Avatar ($p = 0.005$).\\
    \bottomrule[2pt]
    \end{tabular}
}
\caption{Summary of main effects and interactions with post hoc analysis - Affinity and Social Presence.}
\label{tab:result_u}
\end{table}

\section{Discussion}
The result and analysis indicated that realistic avatars got significantly higher ratings in self/other-identification, perceived realism, and social presence, which supports our hypotheses \textbf{H1a}, \textbf{H2a}, and \textbf{H4a} and are in line with previous findings. The post hoc analysis about realistic avatars leading to higher level of social presence is also consistent with the findings of Amadou et al. \cite{NabilaAmadou} and Higgins et al. \cite{photorealismavoidtheuncanny}. Hypothesis \textbf{H3a} about affinity was rejected as participants found realistic MetaHuman avatars to be creepier and less attractive which contradicts some previous research findings. Some previous work have found that the affinity for avatars is unrelated to whether the avatar is realistic or not \cite{IsPhotorealismImportant, zell2015stylize, NabilaAmadou}, while others have found that highly realistic avatars are more attractive \cite{VirtualTwin, RenderMeReal}. However, these studies used virtual humans with random appearances and did not involve realistic digital doubles of real faces that participants were familiar with. That result although being against our initial expectation is in fact aligned with the general observation of people being more sensitive to realistic digital doubles in terms of uncanny valley effect. In our work, although finally no significant interaction effect between avatar familiarity and style was found for any dependent variables, we still can observe a trend from Figure \ref{fig:u_appeal} and \ref{fig:u_eerie} that as the avatar familiarity shifts from self-like to unknown person, the gap in affinity ratings between realistic and cartoon avatars narrows.

\begin{table}[t]
    \centering
    \resizebox{0.99\linewidth}{!}{
    \begin{tabular}{cccccccc}
        \toprule
        \textbf{[H1a]} & \textbf{[H1b]} & \textbf{[H2a]} & \textbf{[H2b]} & \textbf{[H3a]} & \textbf{[H3b]} & \textbf{[H4a]}& \textbf{[H4b]} \\
        \midrule
        \textcolor{green}{A} & \textcolor{green}{A} & \textcolor{green}{A} & \textcolor{green}{A} & \textcolor{red}{R} & \textcolor{green}{A} & \textcolor{green}{A} & \textcolor{red}{R}\\
        \bottomrule
    \end{tabular}
    }
    \caption{Hypotheses revisited. \textcolor{green}{A} means the hypothesis was accepted while \textcolor{red}{R} means the hypothesis was rejected.}
    \label{table:hypotheses}
\end{table}

Hypotheses \textbf{H1b}, \textbf{H2b} and \textbf{H3b} regarding avatar familiarity were supported. When familiar facial features appear on virtual avatars, people tend to notice more details and compare these details with the real person's face in their mind. This leads to stricter and more cautious judgment of avatars with familiar faces and amplifies discomfort and the uncanny valley effect during the comparison process. It is noteworthy that the significant effect of avatar familiarity was limited to dependent variables that included complete visual identity cues. Without body motion capture, participants could not use body movement information for identity recognition. Therefore, avatar familiarity had no significant effect on body movement and behavior identification, which was expected for our experiment as we did not include body motion capture and used a uniform animation for full body instead. Whereas avatar familiarity significantly affected appearance, talking and facial expression identification scales. Especially for realistic digital doubles, when comparing self-like avatar's appearance identification scale with that of interaction partner or unknown person. Furthermore, although avatar familiarity significantly affected talking identification ($p = 0.044$), the Bonferroni pairwise comparison could not detect a significant difference, see Table \ref{tab:result_i}). We believe this is primarily because the avatars’ voices matched their identities perfectly as we recorded the participants’ original voices during facial capture and used them in creating the stimuli. Moreover, there were no severe mismatches between mouth movements and speech ensuring the identity information conveyed by the voice was largely retained. However, the talking identification scores for cartoon avatars were significantly lower than those for realistic avatars. It is to be noted that the simpler ARKit blendshapes were used for cartoon avatars to be in line with their appearance fidelity. The cartoonish shapes and simplistic textures of RPM avatars cannot represent subtle motions (e.g. the dynamic shape changes of lips during speech). We suspect that the contrast between a real person’s voice and the lack of realism in the mouth movements led some participants to not recognize the speech as belonging to a real person.

As for hypotheses \textbf{H2b} and \textbf{H3b}, avatars of unknown persons received significantly higher scores in perceived realism and affinity. Interestingly, the interaction partner's avatar was rated higher than the unknown person's avatar in facial expression identification, overall realism, and facial expression realism. Post hoc tests also indicated that regardless of which avatar ranked the highest, the self-like avatar consistently rated significantly lower than the top-ranked avatar. Regarding hypothesis \textbf{H4b}, there was no statistically significant effect of avatar familiarity on social presence. However, we can observe in Figure \ref{fig:sp} that for realistic digital doubles, the social presence score demonstrates a slight trend where social presence increases as familiarity decreases which is in line with our expectations.

Beyond differences in perceiving familiar versus unfamiliar avatars, the contrast between self-like and interaction partners' avatars is particularly notable. Our experiment showed that participants were highly critical of their own virtual faces. Several noted in qualitative feedback that their avatars resembled them only slightly and felt unsettling. In contrast, they were more lenient when evaluating their partner’s avatar, especially regarding identity recognition and affinity. This aligns with findings by Pakanen et al.~\cite{nicetomeetyou}, who observed that \textit{people preferred to see the other user as photorealistic avatars}. It also follows results from \cite{Familiarity_CAMPBELL2020107415}, which reported the strongest neural responses for one’s own face, followed by a friend’s, and the weakest for a stranger’s— suggesting that personal familiarity amplifies identity-specific processing. Together, these findings indicate people are more sensitive to their own facial representations than to those of familiar others, underscoring the perceptual challenges posed by high-fidelity digital doubles and the importance of their design in shaping user experience.

\subsection{Limitation and Future Work}
Although Unreal Engine MetaHuman pipeline supports the generation of realistic digital doubles, there were several limitations. The LiveLink Face app enabled quick and convenient facial animation capture, however its accuracy was highly dependent on environmental conditions. Proper lighting and close camera proximity were essential; poor lighting, distance from the camera, or rapid head movements (e.g., quick shakes) often led to loss of detail or erroneous data. Such instability introduced abnormal jitters in some participants’ MetaHuman avatar animations, requiring manual corrections via control rig adjustments. These artifacts may have caused certain facial movements to appear unnatural, potentially lowering avatar identification and affinity scores. High-fidelity self-like avatars are especially vulnerable to scrutiny, where even minor anomalies in appearance or motion can lead to discomfort. Using a head rig could mitigate this by maintaining consistent face-camera alignment, but we opted against it to preserve natural head motion because our setup did not include full-body capture. Future work could incorporate simultaneous full-body and facial motion capture to address this limitation. Ideally, a digital double should replicate not only facial features but also an individual’s characteristic body movements. For participants who use expressive upper-body gestures (e.g., shrugging, hand motions), capturing and reproducing these motions would enhance the realism and reliability of such experiments.

Avatar personalization in our experiment was constrained by the available assets in the MH and RPM editors. While we replicated participants’ facial features as closely as possible, exact matches for hairstyles and clothing were not always achievable. For MetaHuman avatars, facial textures often lacked fine-grained detail, which may be negligible for most participants but can be significant for those with distinctive features such as moles, birthmarks, or tattoos. Additionally, RPM avatars offered limited ethnic diversity, potentially affecting the perceived representativeness of these digital doubles for some users. The experimentation with Unreal Engine MetaHumans pipeline shows that although it is possible to generate realistic digital doubles in a reasonable amount of time with consumer grade hardware and software, truly realistic digital doubles still require a lot of manual and artistic effort and high-end appearance and performance capture devices.

A key limitation of our study is the small sample size of 22 participants aged 18 to 34. The controlled lab setup required participants to attend in pairs with real-life interaction partners, which, along with the time-intensive post-processing for avatar creation, constrained recruitment. Future work should include a broader age range to improve generalization and explore whether different age groups perceive digital doubles differently. Additionally, our participant pairs were limited to classmates and co-workers. Including a wider variety of relationships (e.g., family members, short- and long-term friends) could better capture the spectrum of familiarity and strengthen future studies.

Our experiment did not implement real-time interaction with the avatars to keep high quality appearance and animation. Future work can include virtual social environments for a similar experiment in VR where participants would drive their digital doubles in real-time.

\section{Conclusion}
The evidence in our work demonstrates that humans are sensitive and critical of their own avatars and even more so when the avatar is created with a realistic appearance as opposed to a stylized alternative. Our findings reveal that when avatars possess a higher degree of appearance realism, users tend to experience a greater sense of identification, perceived realism, and social presence. This suggests that the more life-like an avatar appears, the stronger the connection users feel towards it. However, an intriguing counterpoint emerged when the avatars featured familiar faces, particularly those with highly realistic appearances. In such cases, the level of identification, perceived realism, and affinity diminished. Participants, while recognizing and identifying their digital doubles as representations of themselves, often expressed dissatisfaction with these avatars. This dissatisfaction was especially pronounced when the avatars were highly realistic. Despite this, participants displayed a more lenient and accepting attitude towards digital doubles representing acquaintances or unknown individuals. In these instances, they were less critical and more forgiving which indicates a complex relationship between familiarity, realism, and user acceptance of avatars. We believe the insights and evidence that this work provides will spark new discussions and research endeavors on how best we can employ digital human avatars. While beneficial for certain applications, digital doubles could become channels for privacy breaches and identity theft. Danaher et al. \cite{Ethics_Danaher2024} presents a minimally viable permissible principle (MVPP) framework that could potentially be adopted when creating digital doubles in order to ensure the proper ethical and privacy considerations.

\bibliographystyle{ACM-Reference-Format}
\bibliography{bibtex}


\end{document}